\documentclass[journal, letterbox]{IEEEtran}
\usepackage{amsmath,amsfonts}

\usepackage{algorithmic}
\usepackage{algorithm}
\usepackage{array}
\usepackage{textcomp}
\usepackage{stfloats}
\usepackage{url}
\usepackage{verbatim}
\usepackage{graphicx}
\usepackage{cite}
\usepackage{xspace}
\usepackage{xcolor}
\usepackage{listings}
\usepackage{listings}
\usepackage{inconsolata} 
\usepackage[most]{tcolorbox}

\usepackage{subcaption}

\usepackage[export]{adjustbox} 
\usepackage{tikz}
\newcommand{\blackcircle}[1]{%
  \tikz[baseline=(char.base)]{
    \node[shape=circle, fill=black, text=white, inner sep=0pt] (char) {\sffamily\footnotesize \textbf{#1}};
  }%
}

\newcommand{\us}{US\xspace}
\newcommand{\llmus}{\textit{LLM US}\xspace}
\newcommand{\stuus}{\textit{Students US}\xspace}
\newcommand{\gtus}{\textit{GT US}\xspace}

\begin{document}

\title{Can LLMs Generate User Stories \\ and Assess Their Quality?}

\author{Giovanni Quattrocchi, Liliana Pasquale, Paola Spoletini, and Luciano Baresi
\thanks{L. Baresi and G. Quattrocchi are with Politecnico di Milano, Dipartimento di Elettronica, Informazione e Bioingegneria, Milan, Italy (email: \{name.surname@polimi.it\}). Liliana Pasquale is with University College Dublin (email: liliana.pasquale@ucd.ie). Paola Spoletini is with Kennesaw State University (email: pspoleti@kennesaw.edu).}}

\markboth{IEEE Transactions on Software Engineering}%
{IEEE Transactions on Software Engineering}

\maketitle

\begin{abstract}
Requirements elicitation is still one of the most challenging activities of the requirements engineering process due to the difficulty requirements analysts face in understanding and translating complex needs into concrete requirements. In addition, specifying high-quality requirements is crucial, as it can directly impact the quality of the software to be developed. Although automated tools allow for assessing the syntactic quality of requirements, evaluating semantic metrics (e.g., language clarity, internal consistency) remains a manual and time-consuming activity.
This paper explores how LLMs can help automate requirements elicitation within agile frameworks, where requirements are defined as user stories (\us). We used 10 state-of-the-art LLMs to investigate their ability to generate \us automatically by emulating customer interviews.  We evaluated the quality of \us generated by LLMs, comparing it with the quality of \us generated by humans (domain experts and students). 
We also explored whether and how LLMs can be used to automatically evaluate the semantic quality of \us. Our results indicate that LLMs can generate \us similar to humans in terms of coverage and stylistic quality, but exhibit lower diversity and creativity.  Although LLM-generated \us are generally comparable in quality to those created by humans, they tend to meet the acceptance quality criteria less frequently, regardless of the scale of the LLM model. Finally, LLMs can reliably assess the semantic quality of \us when provided with clear evaluation criteria and have the potential to reduce human effort in large-scale assessments.
\end{abstract}


\section{Introduction}
\label{sec:intro}

Requirements elicitation is still a challenging activity of the requirements engineering process due to the difficulty requirements analysts face in understanding and translating complex needs into concrete requirements. In addition, the quality of the requirements specification directly impacts the speed of the development iterations~\cite{Dalpiaz.2019, Lucassen.REJ.2016} and the quality of the software to be developed~\cite{Amna.2022}. 
These problems are exacerbated by the transition to agile system development at scale~\cite{Inayat.2015}, where the distributed nature of teams and the demands on long-term maintenance make the need for high-quality requirements even more pressing~\cite{Kasauli.2021}. Although frameworks~\cite{Lucassen.RE.2015} and automated tools~\cite{Lucassen.REJ.2016} exist to evaluate the quality of requirements, assessing semantic metrics (e.g., language clarity, internal consistency) remains a manual and time-consuming activity.

Large Language Models (LLMs)~\cite{BERT,Radford, LLMs}  demonstrated high-level performance in natural language processing tasks, and their use is becoming increasingly prominent. LLMs are pre-trained on an extensive number of natural language and source code documents and have been applied successfully to automate various software engineering tasks~\cite{Fan.2023}, such as
code generation~\cite{Du.ICSE.2024,Ahmed.2024}, program repair~\cite{Jin.FSE.2023,Xia.ICSE.2023,Wei.FSE.2023}, and
comprehension~\cite{Nam.ICSE.2024}.  As system requirements are usually specified in natural language,  LLMs also have the potential to automate requirements engineering activities~\cite{Arora.2024}, particularly requirements elicitation.  Moreover, since requirements quality assessment is usually based on textual analysis, LLMs could also support the evaluation of requirements quality.

In this paper, we explore the capability of LLMs to automate requirements elicitation within agile frameworks, where requirements are defined as user stories (US). We also investigate whether and how LLMs can be used to automatically evaluate the semantic quality of \us. We used 10 state-of-the-art LLMs to emulate a real-world interview-based requirements elicitation process, where an analyst conducts an interview with a customer to gather, interpret and formalize \us. For this purpose, we replicated the experiment conducted by Ferrari et al.~\cite{Ferrari2022RequirementsElicitation}, where 30 students were asked to impersonate the requirements analyst to conduct interviews with a fictitious customer to elicit \us. The \us generated by the students were compared with a set of \us generated by domain experts. Unlike the original experiment, in this paper, for each LLM, we used 30 instances as requirements analysts conducting the interview, and 1 instance impersonated the customer. We asked each instance acting as analyst to generate a maximum of 50 user stories, producing a total of 13958 user stories. We evaluated how well these LLM-generated \us covered those generated by students and domain experts in the original experiment and also estimated their diversity and detectability. 

To investigate the ability of LLMs to evaluate the semantic qualities of \us, 3 authors manually annotated a smaller curated dataset of 153 \us generated by LLMs, students, and domain experts (50, 50, and 53, respectively) using the semantic quality metrics proposed in the Quality User Story (QUS) framework~\cite{Lucassen.RE.2015}. 
We also curated a codebook that provides guidance and examples on how the evaluation process of semantic qualities should be conducted.
Then we asked the LLMs to evaluate the semantic quality of the \us in the curated dataset and provided different levels of guidance based on the codebook.  We evaluated the performance of the different LLMs comparing their estimated semantic quality metrics with the assessments performed by the manual annotators.  
Finally, we evaluated the quality of LLM-generated \us using an automated tool available in previous work (AQUSA~\cite{Lucassen.RE.2015}) for the syntactic metrics and the best performing LLM to evaluate the semantic metrics. 

Our results indicate that although LLMs can generate US similar to human experts in
terms of coverage and stylistic quality, they exhibit lower
diversity and creativity. The most recent models also show reduced detectability, indicating
increased human-likeness in stylistic features.
All LLMs, particularly Claude 3 Opus, could reliably assess US semantic quality when
equipped with clear evaluation criteria (complete codebook) and can potentially reduce the human effort to evaluate a large set of \us.
Although LLMs can generate \us that have a high average syntactic and semantic quality, they tend to produce fewer \us that pass the acceptance criteria compared to those generated by humans, regardless of the size of the model. Common defects include excessive conjunctions and low feature specificity and clarity of the rationale, which reveal the limited capacity of LLMs to contextualize \us. In summary,  LLMs can be used to improve the structural and syntactic quality of human-generated \us. They could support requirements analysts in the generation of an initial set of \us, still requiring human oversight to increase their diversity, specificity, and clarity of rationale.

The rest of the paper is organized as follows. Section~\ref{sec:methodology} presents the replicated experiment on user story generation. Section~\ref{sec:design} describes the study design and research questions. Section~\ref{sec:preparation} outlines the generation and assessment of user stories. Section~\ref{sec:comparison} compares LLM-generated stories with those written by experts. Section~\ref{sec:assessment} investigates the ability of LLMs to evaluate the quality of user stories. Section~\ref{sec:quality} reports on the quality of the stories generated by different LLMs. Section~\ref{sec:threats} discusses threats to validity. Section~\ref{sec:related} reviews related work. Finally, Section~\ref{sec:conclusions} concludes the paper.

\section{Replicated Experiment on US Generation}
\label{sec:methodology}
To evaluate the ability of LLMs to generate \us, we replicated the experiment conducted by Ferrari et al.~\cite{Ferrari2022RequirementsElicitation} using LLMs.
In the original experiment, students participated in an Interview-Based Learning (IBL) process, emulating a real-world requirements elicitation process that allowed them to gather, interpret, and formalize \us.
Each student acted as an analyst and interacted with a fictitious customer, impersonated by one of the study authors.
The \us generation process involved the four steps.

\textbf{Preparation.} Students, acting as analysts, received a brief overview of the target application (a summer camp management system) and were asked to prepare questions for customer interviews to elicit product requirements. The customer was equipped with $53$ \us used as ground truth (\gtus). These \us were part of a widely used existing dataset\footnote{These \us are available at \url{https://zenodo.org/records/13880060} in file \textit{g12-camperplus.txt}.} and served as an initial knowledge base that allowed the customer to answer the students' questions consistently and accurately.

\textbf{Interview.} Each student conducted a 15-minute interview with the customer, during which they asked both their prepared questions and spontaneous follow-up questions. The customer's responses were based on the provided \gtus, which were not disclosed to the students.  The students recorded the interviews and took detailed notes for later analysis. 

\textbf{Follow-up Interview.} After the initial interview, each student reviewed their notes to identify gaps and generate additional questions for a follow-up interview. Then, a second 15-minute interview was conducted, where each student posed their newly generated questions to the customer to refine and expand their understanding of the requirements. 

\textbf{Requirements Elicitation.} Using the information collected from both interviews, each student created between 50 and 60 \us for the system. This range was chosen to roughly match the size of \gtus, allowing a comparative analysis of the generated data. 

Overall, the final dataset contained $1530$ \us produced by $30$ different students (\stuus).

\section{Study Design}
\label{sec:design}
To evaluate the abilities of LLMs to generate and assess \us, our study addresses three research questions (RQs).

\vspace*{2mm}
\noindent
\textbf{RQ1}:  Can LLMs generate \us similar to those elicited by human experts? 

\noindent
\textbf{RQ2}: Can LLMs be used to evaluate the quality of \us?

\noindent
\textbf{RQ3}: What is the quality of the \us generated by LLMs?
\vspace*{2mm}

RQ1 studies whether LLMs can generate \us that have content similar to that elicited by human experts, thus understanding to what extent LLMs can replicate human expertise in requirements elicitation.
RQ2 focuses on the potential of LLMs to evaluate the quality of \us, examining how effectively these models can act as evaluators and how their assessments align with human evaluations. RQ3 aims to measure the quality of the \us generated by LLMs, analyzing quantitative (syntactic) and qualitative (semantic) metrics proposed in the QUS framework~\cite{Lucassen.RE.2015}. Syntactic metrics consider the structure of the user story (e.g., if all required parts are present). Semantic qualities consider the meaning of user stories and their relationships with other user stories. 

\begin{figure*}
    \centering
    \includegraphics[width=0.85\linewidth]{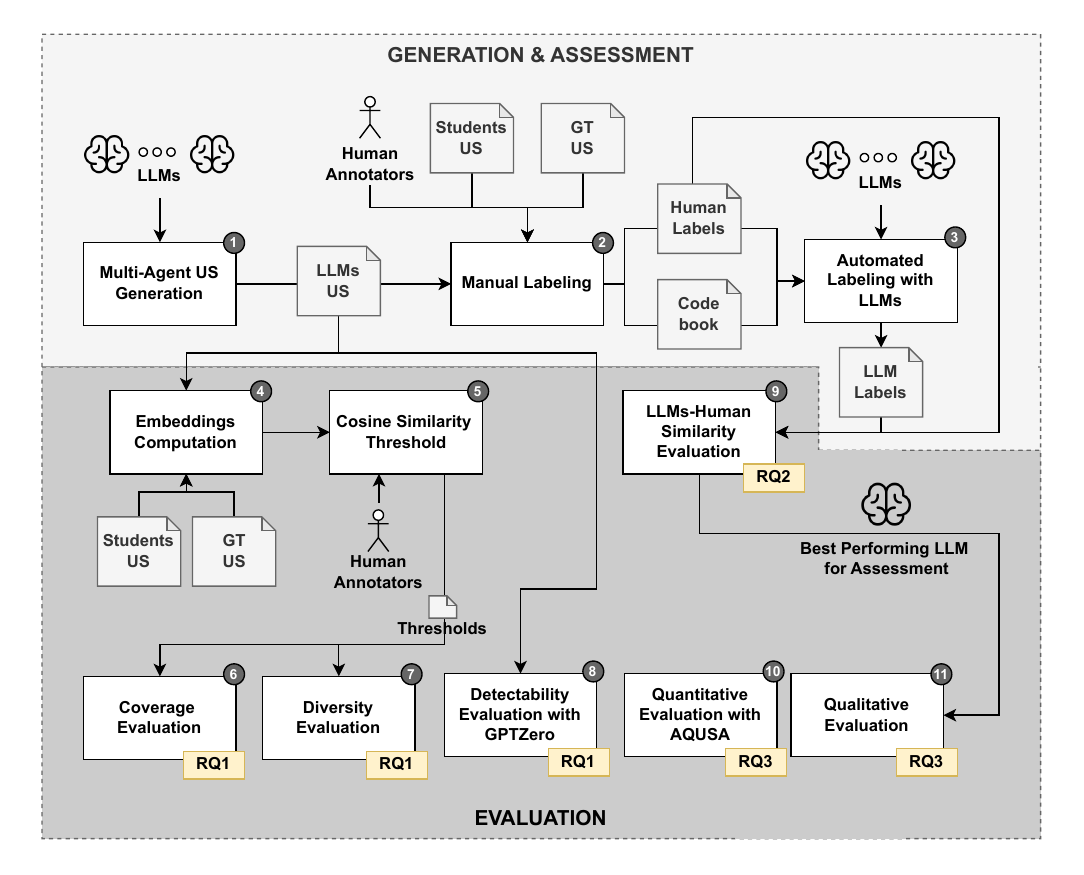}
    \caption{An overview of our study.}
    \label{fig:process}
\end{figure*}

As shown in Fig.~\ref{fig:process},  we divide our study into two phases (Generation \& Assessment and Evaluation). Generation \& Assessment aims to produce the artifacts necessary to answer the research questions: i) a dataset of \us generated by LLMs (\llmus), and ii) a set of labels (LLM Labels) generated by the LLMs when asked to assess a set of \us using qualitative metrics. The upper part of Figure~\ref{fig:process} shows the main steps of this phase (steps \blackcircle{1}-\blackcircle{3}). Evaluation aims to evaluate generated artifacts to answer the RQs.  To answer RQ1, we assess whether the \llmus cover the content of the user stories present in the ground truth (step \blackcircle{6}) and how diverse the \llmus are among one another (step \blackcircle{7}). We also evaluate whether it is possible to detect the user stories generated by LLMs (step \blackcircle{8}).
To answer RQ2, we assess the capabilities of LLMs to evaluate user stories along semantic metrics (step \blackcircle{9}), comparing their labeling with that performed by humans. We do not consider syntactic metrics since an automated tool (AQUSA~\cite{Lucassen.REJ.2016}) is already publicly available for this purpose and achieves high precision and recall.
To answer RQ3, we evaluate the quality of \llmus along the syntactic metrics using AQUSA and the semantic metrics using the LLM that exhibits the best performance when answering RQ2.

\section{User Story Generation and Assessment}
\label{sec:preparation}
\subsection{Multi-Agent \us Generation}
\label{sec:replicatedstudy}

We generated \us using LLMs by replicating the study described in Section~\ref{sec:design} (step \blackcircle{1}).
We implemented the IBL process using LLM as agents acting as analysts (instead of students) and customers (instead of human experts). To our knowledge, this work is the first to replicate the IBL process using a multi-agent LLM architecture. 

We developed the multi-agent architecture using the best-performing LLMs available at the time we started this work, namely \textit{Claude Sonnet 3.5}, 
\textit{Claude Sonnet 3}, 
\textit{Claude Opus 3}, 
\textit{Grok Beta}, 
\textit{GPT-4}, 
\textit{GPT-3.5 Turbo}, 
\textit{LLaMA 3.1 8B}, 
\textit{LLaMA 3.1 405B}, 
\textit{Gemini 1.5 Flash}, 
\textit{Gemini 1.5 Pro}. 
We instantiated each LLM model multiple times to replace the $30$ students and the customer, ensuring consistency in each run. We did not fine-tune any model,  each LLM instance operated in an independent context configured witt the default  sampling temperature. We replicated the IBL process as follows.

\textbf{Preparation.}  We provided the $30$ LLM instances that replaced the students (Analyst LLM) with the same brief description of the app provided to the students in the original experiment. Meanwhile, the additional LLM instance acting as the customer (Customer LLM) received the same \gtus that were originally given to the human expert. 

\textbf{Interview.} Each Analyst LLM engaged with Customer LLM by asking a set of questions. The Customer LLM answered such questions with the \gtus in its context to answer as coherent as possible with the given domain.

\textbf{Follow-up Interview.} Based on the new knowledge acquired, each Analyst LLM generated a new set of questions for further clarification.

\textbf{Requirements Elicitation.} After the second interview, we asked each Analyst to generate $50$ \us, matching the size of the dataset created in the original experiment to enable comparative analysis. 

At the end of requirements elicitation, we obtained an \llmus dataset containing 13958 \us, that is, approximately $50$ \us generated by each of the $30$ LLM agents, instantiated from $10$ different LLM models. 

\subsection{Manual labeling}
\label{sec:manuallabeling}

To establish the ground truth to assess whether LLMs can be used to assess the quality of \us (RQ2), 3 authors (human annotators) independently assessed the semantic qualities of a dataset of 153 \us, which included 53 \gtus, 50 randomly selected \us created by students, and 50 randomly selected \us generated by an LLM (GPT-3.5) (step \blackcircle{2}). The human annotators conducted the labeling process blindly, which means that they were unaware of the source of each \us.

The human annotators independently labeled the \us according to the following semantic metrics proposed in the QUS framework~\cite{Lucassen.RE.2015}. We focused on qualities applied to a single US, not the entire set. Avoiding qualities based on inter-story dependencies allowed us to assess the capabilities of LLMs more directly, yielding clear and actionable insights into their strengths and weaknesses. We measured \textbf{conceptual soundness} through \textbf{Feature Specificity (FS)}, which measures the specificity with which the \us describes a feature, and highlights how specific and direct the explanation is, and \textbf{Rationale Clarity (RC)}, which measures how clearly the reason or justification behind the need for a feature is articulated in a \us. It assesses the clarity with which the user's goal or problem is explained.
We considered \textbf{Problem-Oriented (PO)} to measure how clearly the problem is specified without implying solutions. Finally, we addressed \textbf{unambiguity} by means of \textbf{Language Clarity (LC)}, which measures the clarity and precision of the language used, and \textbf{Internal Consistency (IC)} to measure whether \us is internally consistent and free of contradictions.

The human annotators evaluated each metric using a three-point scale: non-acceptable (1), acceptable but has margins of improvement (2), or good (3). In addition, they also documented the rules and heuristics they used during the evaluation process. This additional layer of annotation captured implicit decision-making strategies and improved the consistency of the assessment framework.
To ensure consistency and reliability in the manual annotation process, we conducted a multi-stage reconciliation procedure. Initially, two annotators independently labeled the same set of \us according to the predefined criteria. They then compared their labels and resolved any major disagreements through discussion. The resulting reconciled set was further reviewed and refined by a third annotator, allowing for additional verification and harmonization of the labeling strategy.

We systematically analyzed disagreements among annotators to identify \textit{conflicts}, which we define as substantial mismatches in the perceived quality of a user story. Specifically, a conflict occurred when one annotator assigned a score of 1 (non-acceptable) and another assigned either a 2 (acceptable) or a 3 (good) for the same quality criterion on the same user story. Formally, let $S_{A}$ and $S_{B}$ denote the scores assigned by two annotators $A$ and $B$. A disagreement is considered a conflict if:
\begin{equation} 
\label{eq:conflict}
\text{Conflict}(S_A, S_B) = \begin{cases} 1 & \text{if } S_A = 1 \land S_B \in {2,3} \\ 
1 & \text{if }  S_B = 1 \land S_A \in {2,3}\\ 
0 & \text{otherwise} \end{cases} \end{equation}
This approach avoids over-penalizing minor variations (e.g., scoring a \us as 2 vs. 3), and instead targets more substantial discrepancies.

\begin{lstlisting}[caption={Codebook Excerpt}, label={lst:codebook_rc}, float=t    ]
Rationale Clarity (RC)
- The US should clearly justify the actor's rationale.
- To assign score 3, the US should have the rationale explicitly expressed through a subsentence that could start with "so that," "as a" or similar phrases.
- The US should include a rationale or a description of the problem; otherwise, it should be assigned score 1.
- If the rationale is implied through details given in the feature description, score 2 should be assigned.

Examples:
- Score 3: "As a parent, I want to be able to enroll my children so that they can be admitted to camp."
- Score 2: "As the administrator, I would like the system to have a forum area where I can communicate directly with all employees about camp rules or feedback."
- Score 1: "As a camp administrator, I want to be able to create and modify rules that campers and camp workers have to follow."
\end{lstlisting}

At the end of the reconciliation process, the evaluation rules and guidelines formulated by the annotators were compiled into a structured \textit{codebook}, providing a reference framework for assessing \us quality in future analyses.
Listing~\ref{lst:codebook_rc} is an excerpt from the codebook compiled during the reconciliation process, illustrating the evaluation criteria applied to the Rationale Clarity metric. 

The excerpt provides both the evaluation criteria and illustrative examples. The examples highlight how different levels of rationale clarity manifest in practice.
\us rated with a score 3 provide a clear and explicit rationale, making the actor's intent fully transparent, as illustrated in the first example. In contrast, score 2 \us contain an implicit rationale, where the reasoning can be inferred but is not explicitly stated (see the second example). Lastly, score 1 \us lack a rationale, making it difficult to understand the underlying motivation for the requirement (as demonstrated in the third example). 

\subsection{Automated Labeling with LLMs}

To evaluate the ability of LLMs to assess \us, we asked the 10 selected LLMs to evaluate the same 153 \us that had been manually annotated (step \blackcircle{3}). We experimented with three different prompting approaches to analyze the impact of instruction specificity on LLM performance: i) \textit{basic prompt (no codebook)}: a simple prompt explaining the evaluation task, instructing the LLM to assess each \us based on predefined quality criteria, ii) \textit{codebook-augmented prompt}: the same base prompt, supplemented with the full codebook developed during the manual annotation phase, and iii) \textit{partial codebook-augmented prompt}: a refined version of the previous approach, where a more concise summary of the codebook was included to balance informativeness and brevity.
These three prompting strategies allowed us to assess the effect of additional contextual guidance on LLM-based evaluations and compare their evaluations with those performed by human annotators.

\section{Comparing LLM and Expert User Stories}
\label{sec:comparison}

 To evaluate the similarity between LLM- and human-generated \us (RQ1) we used three metrics: i) coverage, ii) diversity, and iii) detectability.

\noindent\textbf{Coverage.} Coverage measures how well a set of \us captures the requirements expressed in the \gtus. Specifically, it quantifies the semantic overlap between the two sets of \us using embedding-based similarity.

To calculate coverage, we first converted all the \us (\gtus, \llmus, \stuus) into a vector representation using OpenAI's \textit{text-embedding-3-small} model (step \blackcircle{4}). For any two sets of \us $X$ and $Y$, we define the coverage of $X$ with respect to $Y$  as the percentage of \us in $Y$ with at least one semantically similar counterpart in $X$. Two \us are considered semantically similar if their \textit{cosine similarity} equals or exceeds a set threshold $c$.

More formally, for each \us $y \in Y$, we compute its cosine similarity (step \blackcircle{5}) with every \us $x \in X$. If at least one similarity value meets or exceeds the threshold $c$, we consider $y$ to ``covered" by $X$. We calculated the coverage score as:

\begin{equation}
\label{eq:coverage}
Coverage(X,Y) = \frac{\text{Number of covered stories in } Y}{\text{Total number of stories in } Y} \times 100
\end{equation}

\noindent where a story $y \in Y$ is considered covered if:
\begin{equation}
\exists x \in X : cosine\_similarity(x,y) \geq c
\end{equation}

We measured cosine similarity within a range of [-1,1], where 0 indicates no similarity, 1 represents identical texts and while -1 texts with opposite meanings. To determine an appropriate threshold for considering two \us as semantically similar, we empirically identified two possible thresholds: 0.8 (less strict) and 0.85 (more conservative). We selected these thresholds based on a human evaluation study designed to calibrate similarity assessments.
 We used three human annotators to assess pairs of \us manually. For each similarity bin (ranging from 0.6 to 0.85 in steps of 0.05), we randomly selected 20 pairs of \us from any set. The annotators independently determined whether the two \us conveyed the same meaning without being informed of the actual similarity score. A pair was considered a match if at least two out of three annotators agreed that the \us were similar. 

Table~\ref{tab:cosine_similarity} summarizes the results of this human evaluation study.
\begin{table}[t]
\centering
\caption{Determining the similarity thresholds.}
\renewcommand{\arraystretch}{1.2}
\begin{tabular}{|c|c|c|c|}
\hline
\textbf{Bin} & \textbf{\% Match} & \textbf{\# Matches} & \textbf{\# Count} \\
\hline
0.6-1  & 28.3\% & 34  & 120 \\
0.65-1 & 34.0\% & 34  & 100 \\
0.7-1  & 42.5\% & 34  & 80  \\
0.75-1 & 53.3\% & 32  & 60  \\
0.8-1  & 72.5\% & 29  & 40  \\
0.85-1 & 95.0\% & 19  & 20  \\
\hline
\end{tabular}
\label{tab:cosine_similarity}
\end{table}
The results show that at a cosine similarity threshold of 0.8, human annotators considered 72.5\%  of the \us pairs similar, making it a reasonable threshold to capture semantic equivalence. At 0.85, the agreement increased to 95.0\%, indicating a stricter but also more reliable threshold. Lower thresholds, such as 0.6 or 0.65, resulted in considerably lower agreement levels, suggesting a higher risk of including semantically dissimilar stories. These insights guided our decision to use 0.8 as the threshold for coverage evaluation and 0.85 for more conservative assessments.
    
This metric provides a quantitative measure of the comprehensiveness with which one set of \us captures the requirements present in another set, making it particularly useful for comparing different approaches to requirements elicitation or evaluating the completeness of requirements specifications.

\textbf{Diversity.}  Diversity measures the extent to which different sources (or generators) ---such as multiple instances of the same LLM or different students--- produce distinct sets of \us. While \textit{coverage} quantifies how well one set captures another, \textit{diversity} captures the semantic variability among different generators of \us. High diversity indicates that the generators produce complementary and varied content, while low diversity suggests a convergence toward similar outputs.
To compute diversity, we build on the coverage metric. Specifically, we define the diversity between two sets of \us $X$ and $Y$ as the complement of the coverage between them:

\begin{equation} 
\label{eq:diversity}
Diversity(X, Y) = 100 - Coverage(X, Y) 
\end{equation}

\noindent where $Coverage(X, Y)$ is defined as in Equation \ref{eq:coverage}, and measures the percentage of stories in $Y$ that are semantically covered by $X$, using $c=0.8$.
To estimate the overall diversity of a group ---such as a cohort of students or multiple instantiations of the same LLM--- we compute pairwise diversity scores across all distinct pairs of \us sets within the group. Let $\mathcal{S} = {S_1, S_2, \dots, S_n}$ be the collection of $n$ \us sets generated by $n$ individuals or instances. The average pairwise diversity within the group is given by:

\begin{equation} 
AvgDiversity(\mathcal{S}) = \frac{2}{n(n-1)} \sum_{1 \leq i < j \leq n} Diversity(S_i, S_j) 
\end{equation}

\noindent where each $Diversity(S_i, S_j)$ is computed using Equation \ref{eq:diversity}.
This formulation captures how varied the \us sets are among different sources. In our study, we apply this metric to compare the diversity among different instances of the same LLM and among \us produced by different students. 

\textbf{Detectability.} To assess the detectability of generated \us, we used GPTZero's API\footnote{\url{https://gptzero.me/technologye v}}, an AI detection tool that uses deep learning techniques to identify AI-generated content. For each LLM, we analyzed 50 randomly selected \us and calculated the average AI probability score.

GPTZero implements an end-to-end deep learning approach, utilizing a sentence-level classifier to determine the probability of AI authorship. The tool is trained on various text datasets, including web content, educational material, and outputs from various LLMs. For each \us, GPTZero returns a probability score indicating the likelihood of AI generation.

We aggregated these individual scores to compute an average AI probability for each LLM, student-generated content, and Ground Truth stories. This methodology provides a quantitative measure of how distinguishable each source's writing patterns are from human-authored content, offering insights into the detectability of AI-generated requirements in software development contexts.

\subsection{Coverage}

\begin{table}[t]
    \centering
    \caption{Coverage Results of LLMs against Ground Truth (Sorted by $c=0.8$)}
    \label{tab:llm_coverage}
    \begin{tabular}{|lcc|}
    \hline
    \textbf{LLM} & \textbf{c=0.8} & \textbf{c=0.85} \\
    \hline
    Claude 3 Sonnet & 96.23 & 88.68 \\
    Claude 3 Opus  & 96.23 & 84.91 \\
    Gemini 1.5 Pro & 90.57 & 77.36 \\
    Llama 3.1 405B & 81.13 & 64.15 \\
    Grok & 81.13 & 43.40 \\
    Claude 3.5 Sonnet & 79.25 & 49.06 \\
    Gemini 1.5 Flash & 77.36 & 50.94 \\
    GPT-4 & 75.47 & 43.40 \\
    GPT-3.5 & 73.58 & 37.74 \\
    Llama 3.1 8B & 58.49 & 16.98 \\
    \hline
    \textit{Total LLM Coverage} & 98.11 & 98.11 \\
    \hline
    \textit{Students Coverage} & 52.83 & 15.09 \\
    \hline
    \end{tabular}
\end{table}

We evaluated the coverage performance of the LLMs and students in step \blackcircle{6}. The results, shown in Table~\ref{tab:llm_coverage}, show that LLMs consistently outperform students in generating \us (\us) that align with the ground truth (\gtus). At both similarity thresholds (0.8 and 0.85), LLMs demonstrate significantly higher coverage, meaning they produce \us more similar to the reference set of validated requirements.

For example, \textit{Claude 3 Sonnet} and \textit{Claude 3 Opus} achieve a top coverage of 96.23\% at the 0.8 threshold, while even the lowest-scoring LLM (\textit{GPT-3.5}) reaches 73.58\%. In contrast, students only reach 52. 83\% at the same threshold. When the threshold is raised to 0.85, the gap widens further: \textit{Claude 3 Sonnet} and \textit{Claude 3 Opus} retain high coverage (88.68\% and 84.91\%), but students drop sharply to just 15.09\%.

This trend suggests that LLMs are highly capable of capturing the \emph{core} requirements contained in the \gtus. Their outputs tend to be syntactically well-formed and semantically consistent with the given set. This is likely due to the general-purpose training on large corpora of structured text and task-oriented patterns. In contrast, students struggle more with structural consistency and may miss some fundamental aspects of requirements expression.

However, these results must be interpreted with care. The ground truth used for evaluation only represents a portion of the full requirement space ---typically, the system's more standard or fundamental needs. High coverage with respect to this reference set is not necessarily indicative of a model's ability to explore the broader, more creative, or domain-specific requirements that stakeholders may expect. The tendency of LLMs to stay close to known patterns may limit their ability to propose novel or unconventional \us. Therefore, while the high overlap with the \gtus shows precision and alignment, it could also reflect a conservative generation strategy that favors familiarity over exploration. In terms of temperature, we used the default one for all the LLMs, which guarantees a balanced tradeoff between creativity and consistency in the generated outputs. Higher values of the temperature, may have lead to lower coverage values favoring the generation of more diverse and nuanced \us.

Moreover, the higher coverage of LLMs might lead to over-reliance on automated tools, potentially reducing stakeholder involvement and creativity in the requirements engineering process. Despite their lower performance in this experiment, human analysts may still contribute more nuanced, contextual, or innovative perspectives that are difficult to evaluate through coverage alone. This hypothesis is supported by the results we obtained for the diversity metric.

\begin{tcolorbox}[colback=gray!5!white, colframe=black!60!black, title=Insights on Coverage Evaluation]
LLMs exhibit strong capabilities in reproducing basic requirements, achieving high coverage against ground truth \us. However, this strength may (or may not) reflect a narrow focus on standard patterns, potentially limiting creativity and completeness in broader requirement spaces.
\end{tcolorbox}

\subsection{Diversity}

\begin{table}[t]
\centering
\caption{Average diversity of LLMs and students}
\label{tab:diversity}
\begin{tabular}{|lc|}
\hline
\textbf{Generator} & \textbf{$AvgDiversity$ (\%)} \\
\hline
Students                  & 98.58 \\
Grok Beta                 & 74.74 \\
LLaMA 3.1 8B              & 73.36 \\
LLaMA 3.1 405B            & 65.67 \\
GPT-4                     & 62.21 \\
Claude 3.5 Sonnet         & 56.08 \\
Claude 3 Opus             & 55.35 \\
Gemini 1.5 Flash          & 52.85 \\
Gemini 1.5 Pro            & 49.38 \\
GPT-3.5                   & 48.38 \\
Claude 3 Sonnet           & 44.23 \\
\hline
\end{tabular}
\end{table}

\begin{figure*}
    \begin{subfigure}{0.18\textwidth} 
    \centering
    \raisebox{0.45\textwidth}{\includegraphics[width=\textwidth]{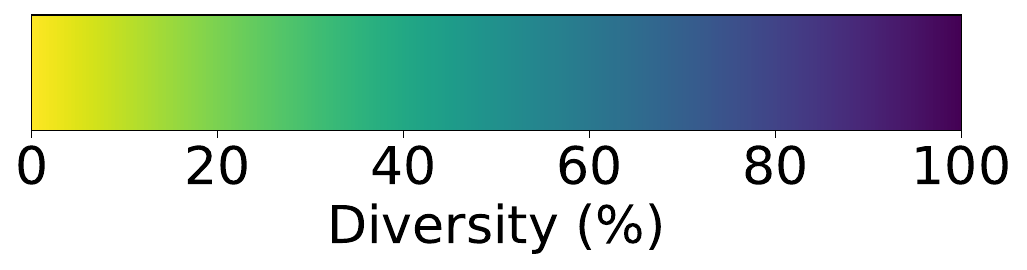}}
    \end{subfigure}
    \hfill
    \begin{subfigure}{0.18\textwidth}
\includegraphics[width=\textwidth]{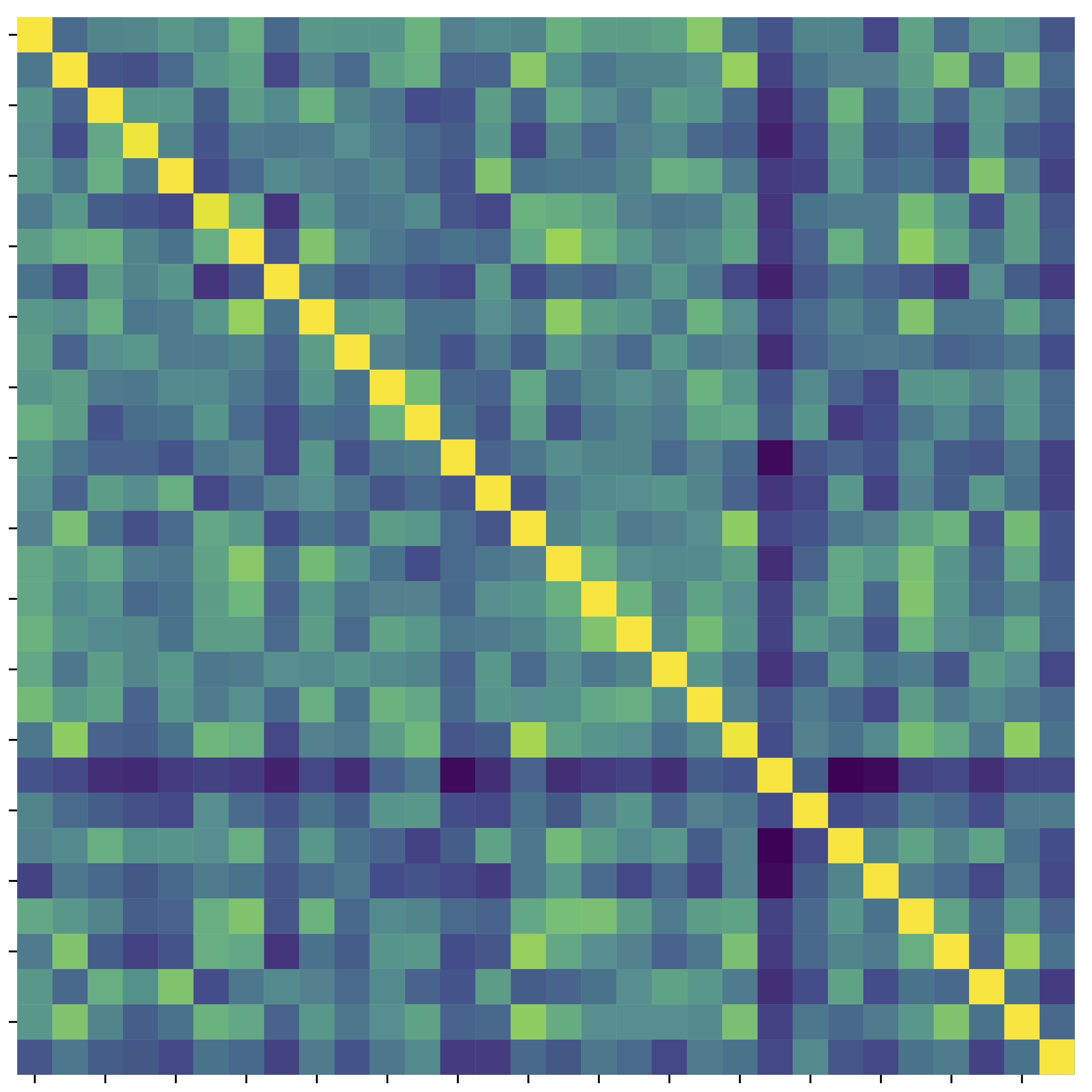}
    \caption{Claude Sonnet 3.5}
    \end{subfigure}
    \hfill
    \begin{subfigure}{0.18\textwidth}
    \includegraphics[width=\textwidth]{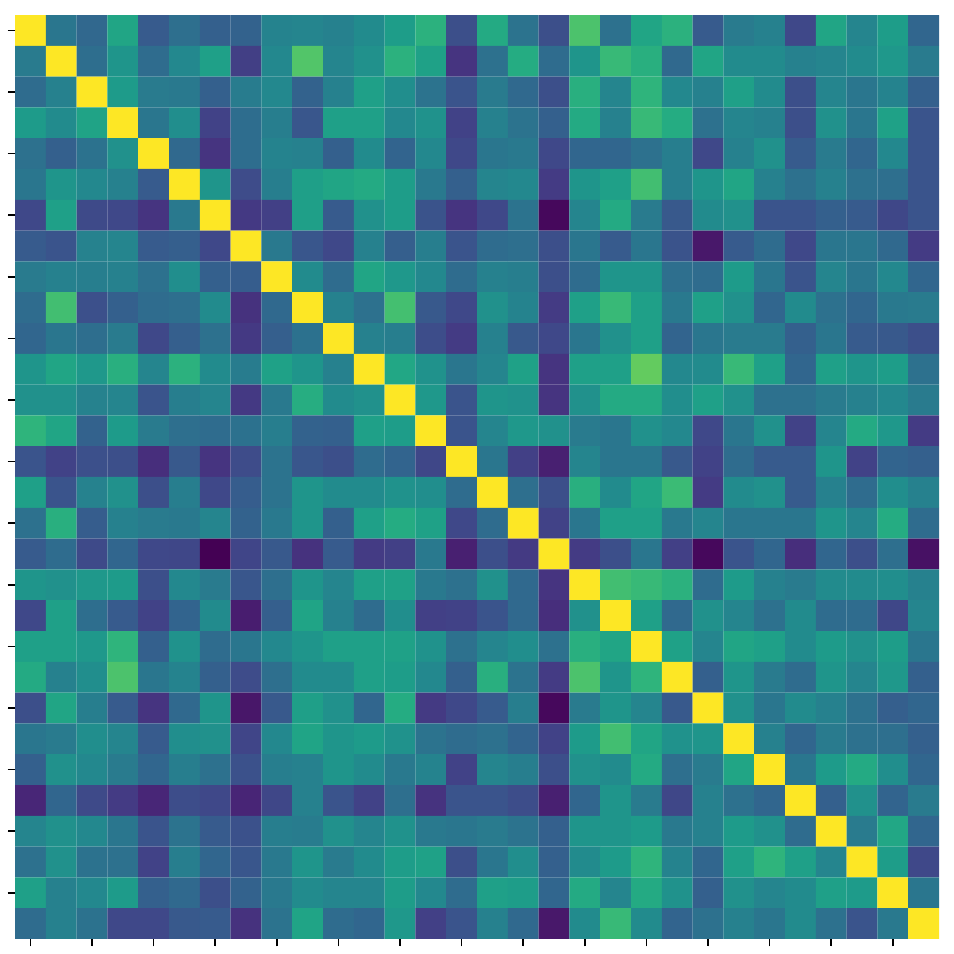}
        \caption{Claude 3 Opus}
    \end{subfigure}
    \hfill
    \begin{subfigure}{0.18\textwidth}
    \includegraphics[width=\textwidth]{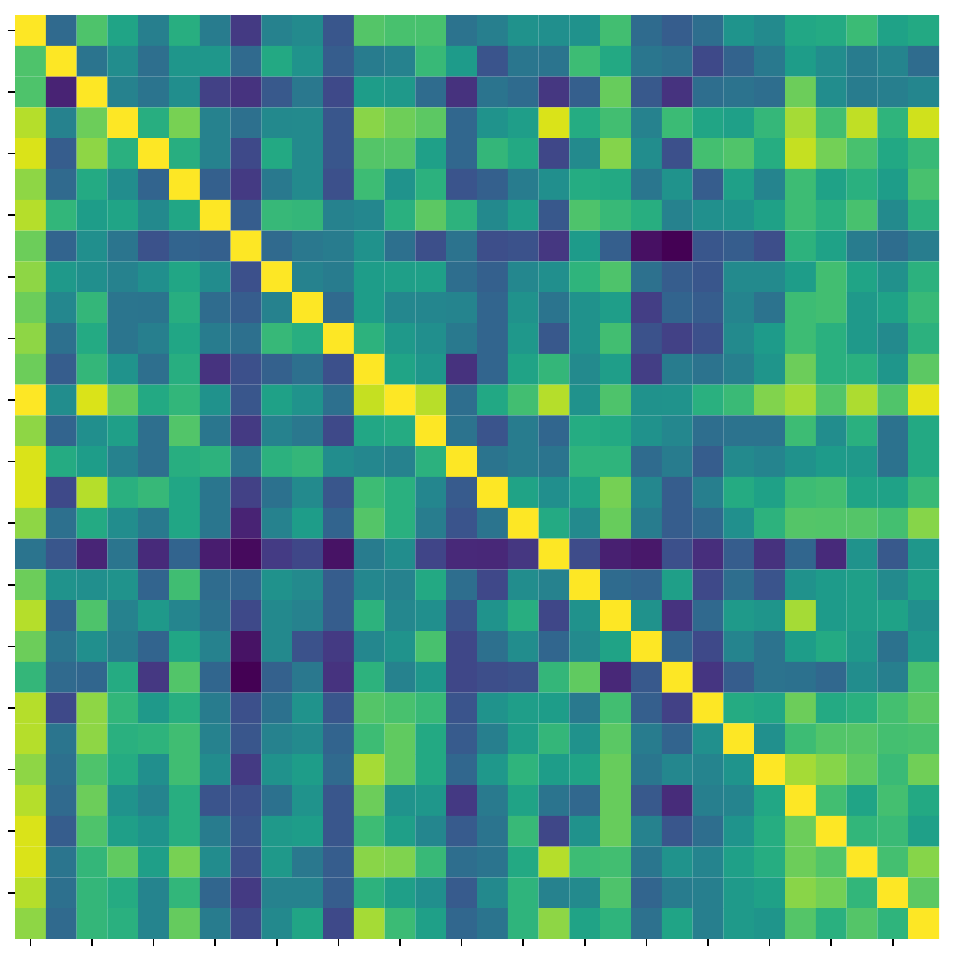}
        \caption{Claude 3 Sonnet}
    \end{subfigure}
    \hfill
    
    \begin{subfigure}{0.18\textwidth}
    \includegraphics[width=\textwidth]{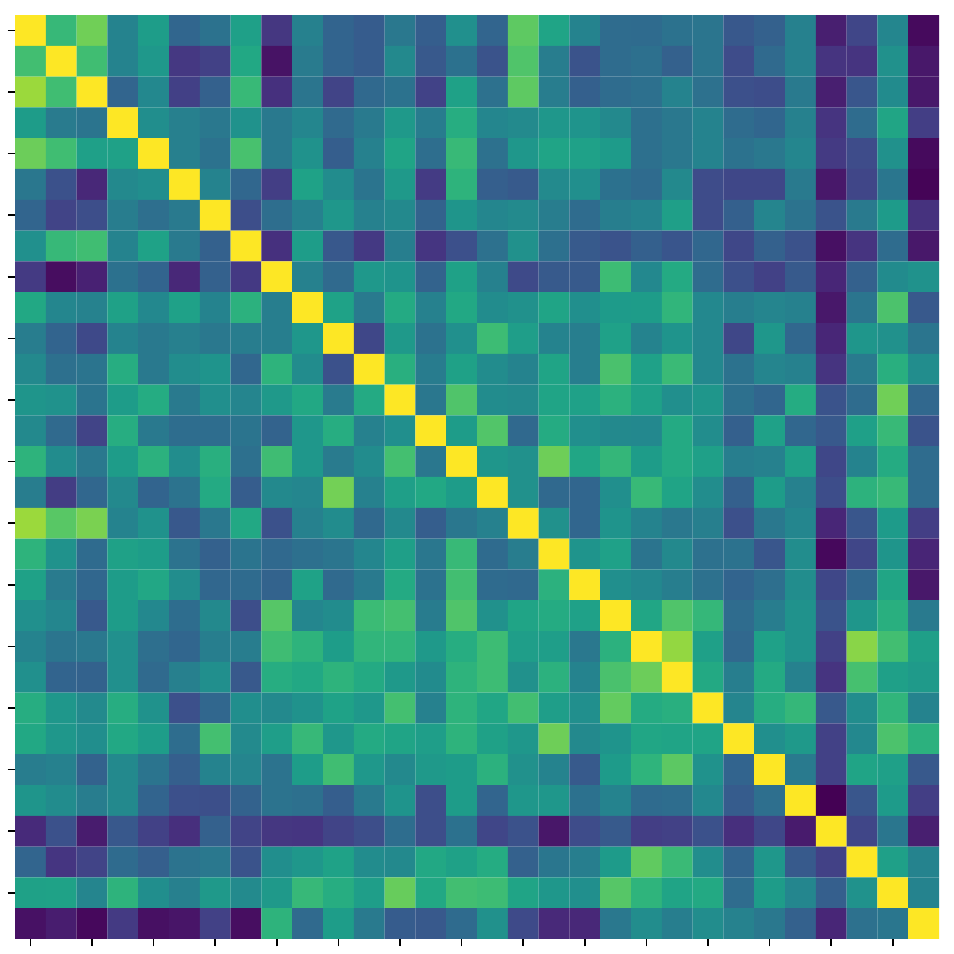}
        \caption{Gemini 1.5 Flash}
    \end{subfigure}
    \hfill
    \begin{subfigure}{0.18\textwidth}
    \includegraphics[width=\textwidth]{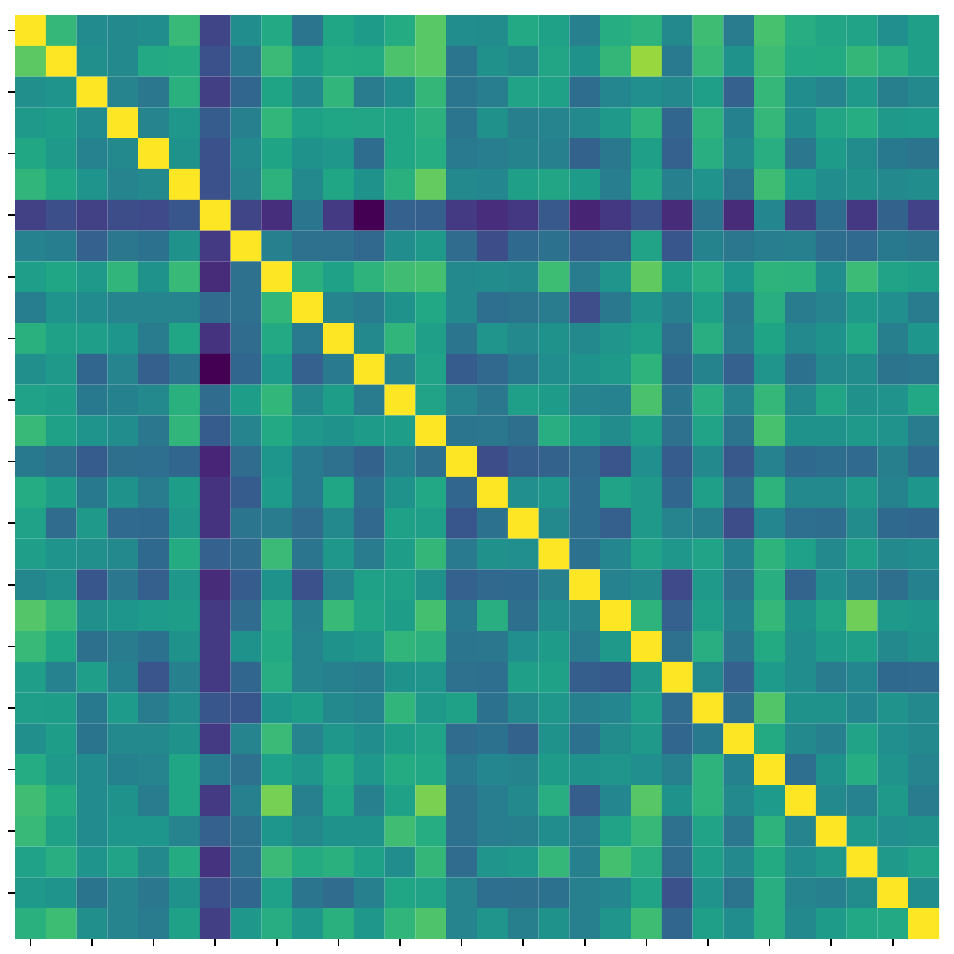}
        \caption{Gemini 1.5 Pro}
    \end{subfigure}
    \hfill
    \begin{subfigure}{0.18\textwidth}
    \includegraphics[width=\textwidth]{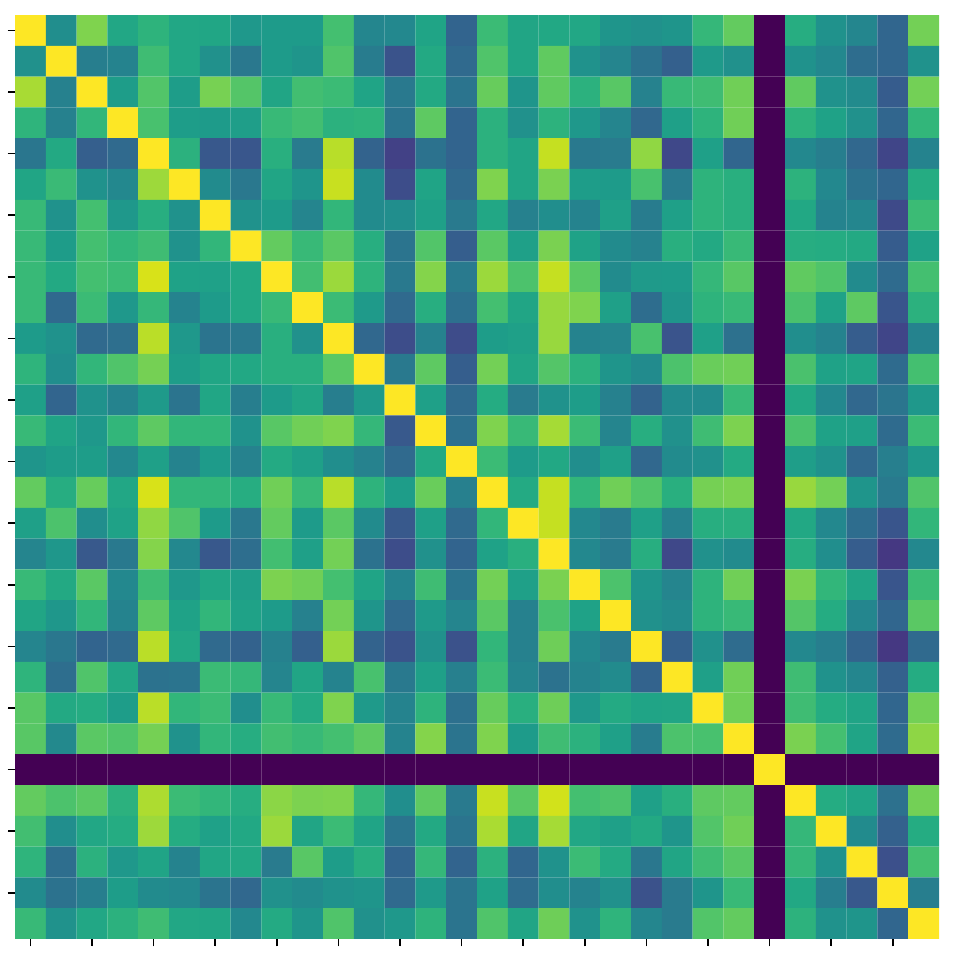}
        \caption{GPT-3.5}
    \end{subfigure}
    \hfill
    \begin{subfigure}{0.18\textwidth}
        \includegraphics[width=\textwidth]{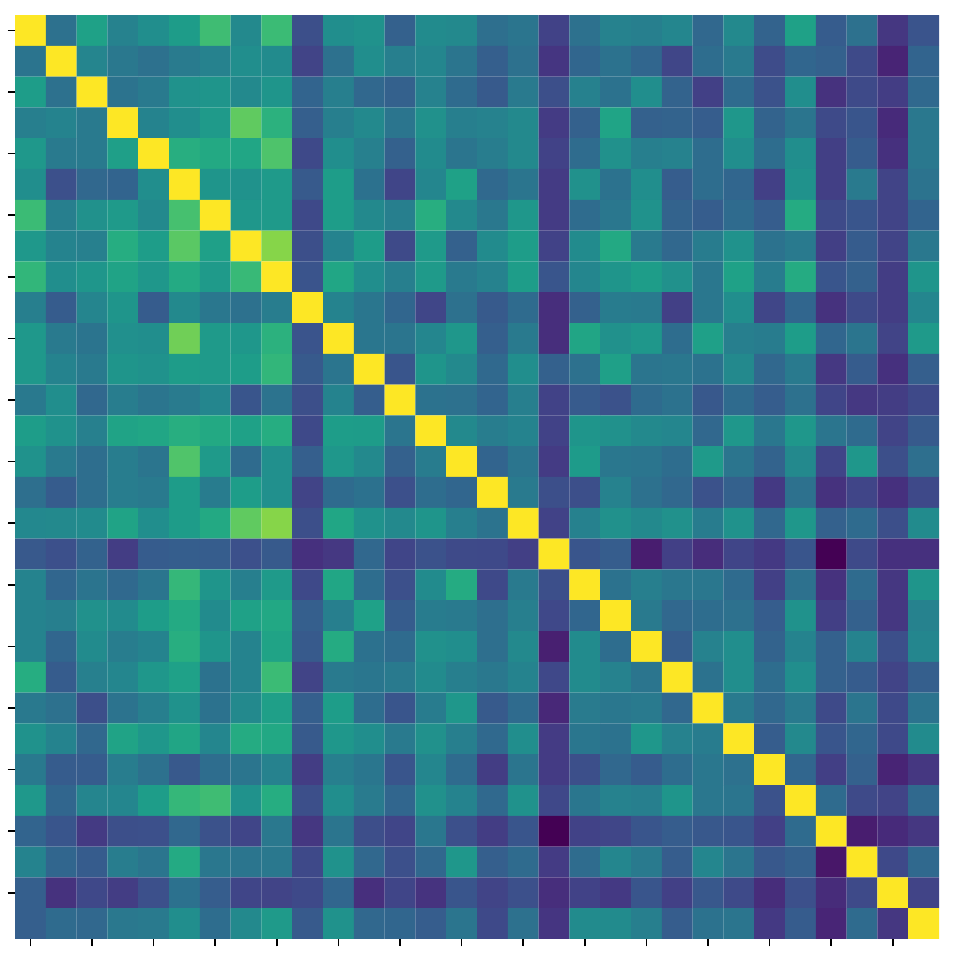}
        \caption{GPT-4}
    \end{subfigure}
    \hfill
    
    \begin{subfigure}{0.18\textwidth}
        \includegraphics[width=\textwidth]{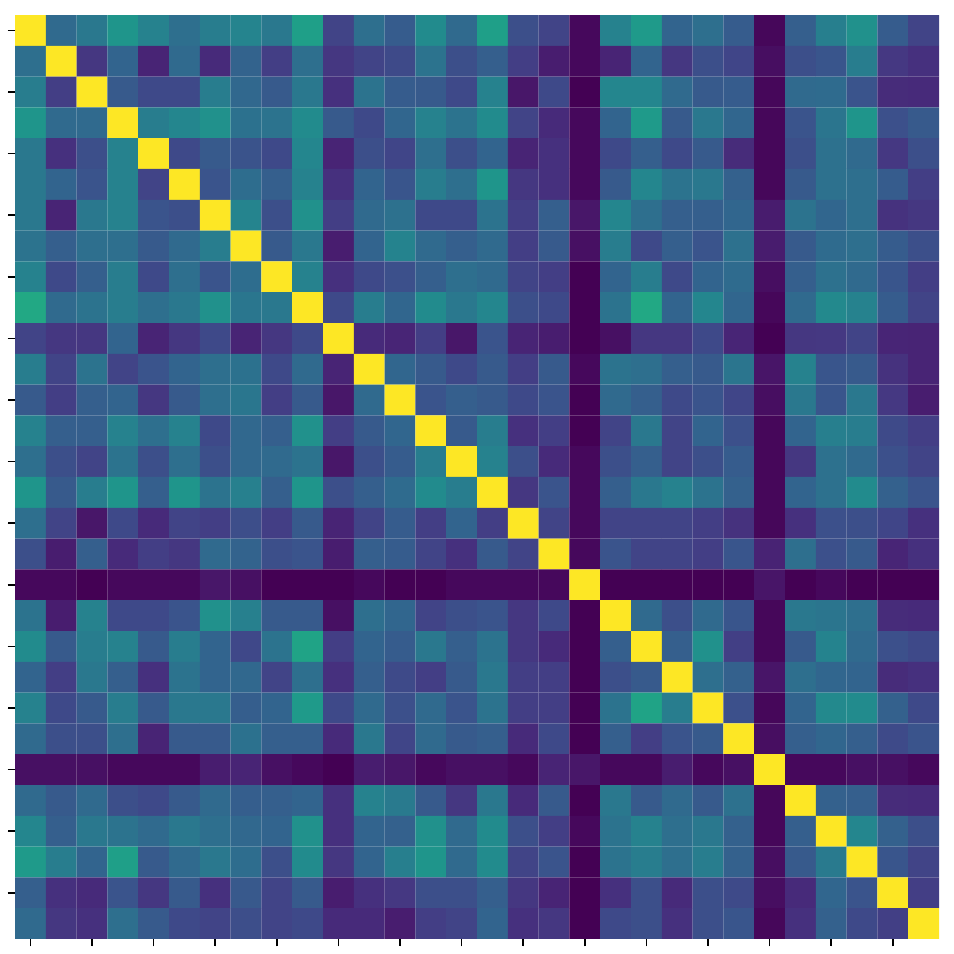}
        \caption{Grok Beta}
    \end{subfigure}
    \hfill
    \begin{subfigure}{0.18\textwidth}
    \includegraphics[width=\textwidth]{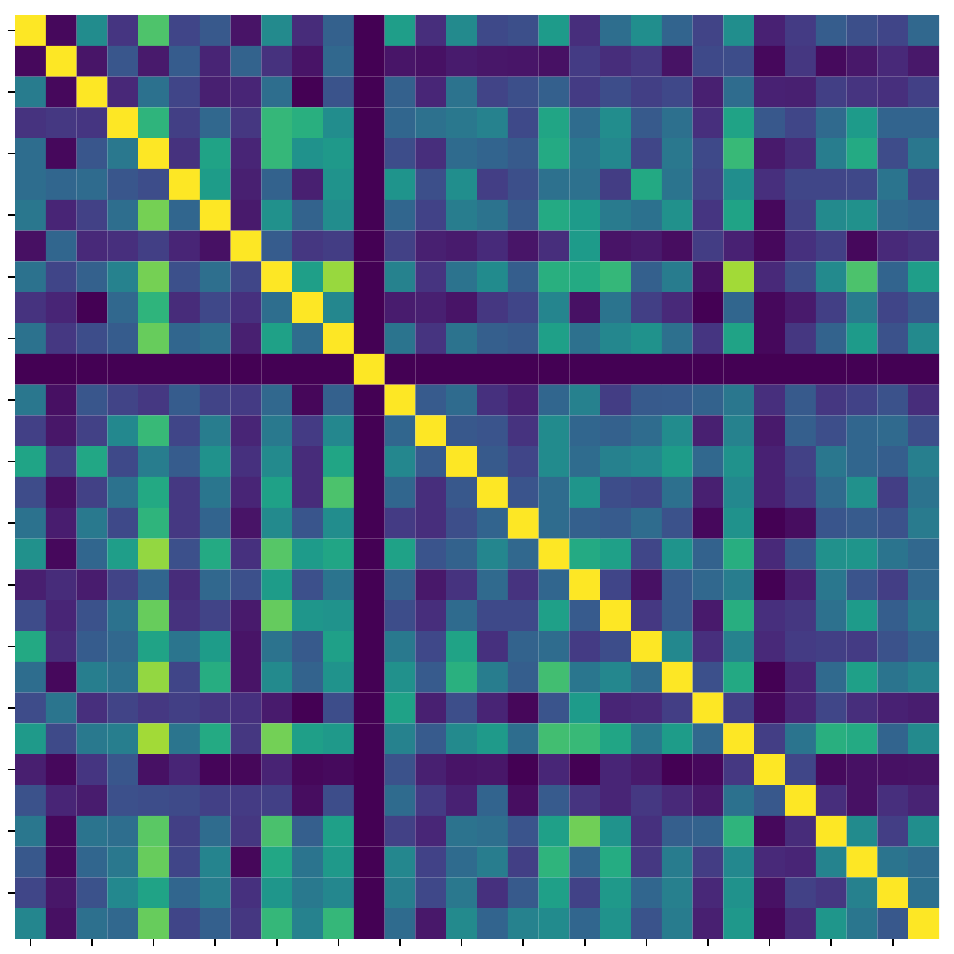}
        \caption{Llama3.1 8B}
    \end{subfigure}
    \hfill
    \begin{subfigure}{0.18\textwidth}
    \includegraphics[width=\textwidth]{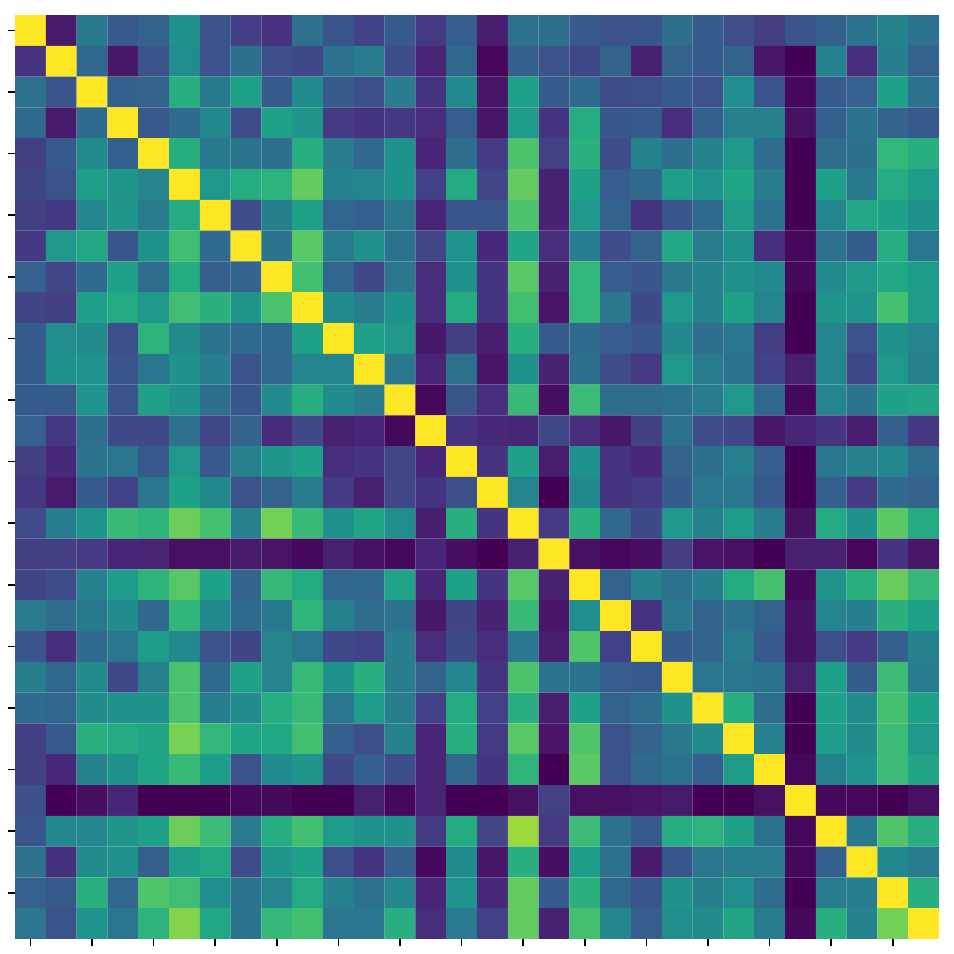}
        \caption{Llama3.1 405B}
    \end{subfigure}
    \hfill
      \begin{subfigure}{0.18\textwidth}
    \includegraphics[width=\textwidth]{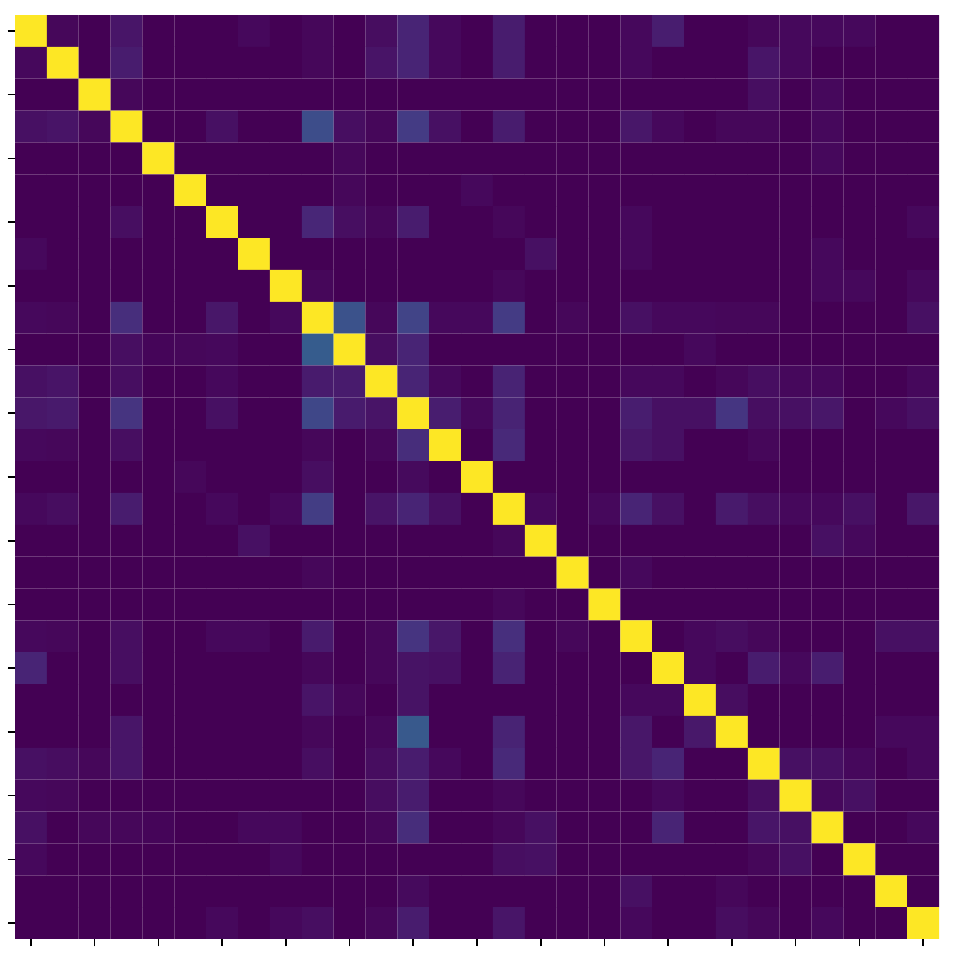}
        \caption{Students}
    \end{subfigure}

    \caption{Diversity across different LLMs and Students}
    \label{fig:diversity}
\end{figure*}

The results for diversity evaluation (step \blackcircle{7}) are summarized in Table~\ref{tab:diversity}, and visualized through heat maps in Figure~\ref{fig:diversity}, where darker colors indicate greater pairwise diversity.

The students exhibit the highest diversity (98.58\%), far surpassing any LLM. This is clearly visible in Figure~\ref{fig:diversity}k, where the off-diagonal elements of the matrix are almost uniformly dark, indicating minimal redundancy. In contrast, many LLMs show brighter horizontal or vertical bands, revealing runs that consistently produce similar outputs. These artifacts point to latent deterministic behavior or inadequate stochasticity in decoding.

This result confirms the intuition that human writers, even when presented with the same task, tend to interpret and express requirements in highly distinct ways. Their variation stems not only from vocabulary and sentence structure, but also from different interpretations of domain concepts, priorities, and expression styles.
Among LLMs, the highest diversity is observed in Grok Beta (74.74\%) and LLaMA 3.1 8B (73.36\%), followed by the larger LLaMA 3.1 405B (65.67\%) and GPT-4 (62.21\%). This suggests that these models introduce more internal variability across independent generations, possibly due to architectural differences, decoding strategies, or higher sensitivity to random sampling noise.
In contrast, models like Claude 3 Sonnet (44.23\%) and GPT-3.5 (48.38\%) produce significantly more homogeneous outputs across different runs. 

The diversity in generated requirements has complex implications. On the one hand, high diversity can be advantageous in the early stages of requirements elicitation, where a wider set of ideas is valuable. It allows broader coverage of the requirements space and may surface alternative perspectives. For example, the behavior of Grok Beta and LLaMA 3.1 8B in this respect may be desirable in contexts where creative variability is important.
On the other hand, excessive diversity might be problematic for standardization or alignment with domain-specific terminology. A more consistent generation (as seen with Claude 3 Sonnet or GPT-3.5) could be preferable in such settings, particularly when the LLM is integrated into structured pipelines. Moreover, diversity must be evaluated jointly with \emph{coverage}: a model that produces diverse but irrelevant stories is not necessarily useful.

Interestingly, larger models do not always produce more diverse outputs. For example, LLaMA 3.1 405B (405 billion parameters) shows lower diversity than its smaller counterpart, LLaMA 3.1 8B. This may indicate that larger models are better at converging toward high-probability, canonical phrasings, effectively reducing internal variability. This observation cautions against assuming that scaling up model size inherently improves the diversity of generated content.

The sharp contrast between student diversity and that of all LLMs underscores the gap in internal variability. Although LLMs may approach human-level performance in fluency, they appear to lack the range of semantic and expressive variation seen in real users. Increasing the temperature parameter beyond the default value can enhance the diversity of LLM outputs, potentially leading to more varied generations.

\begin{tcolorbox}[colback=gray!5!white, colframe=black!60!black, title=Insights on Diversity Evaluation]
Student-authored \us show far greater internal diversity than those generated by LLMs, confirming that humans naturally introduce variation in language, interpretation, and focus. Some LLMs ---such as Grok Beta and LLaMA 3.1 8B--- witness higher diversity levels, which may be beneficial in early-stage ideation or exploratory requirements elicitation. Overall, diversity should be interpreted as a double-edged property: valuable when exploring breadth but potentially limiting when aiming for alignment and stability.
\end{tcolorbox}

\subsection{Detectability}

Table~\ref{tab:detectability} deals with detectability (step \blackcircle{8}), and shows the average AI probability assigned by GPTZero for each source. This metric reflects how likely a text is to be classified as AI-generated, based on surface-level and stylometric features, such as sentence structure, lexical diversity, burstiness, and perplexity patterns.

\begin{table}[t]
\centering
\caption{Results on GPTZero AI Probability Detection.}
\label{tab:detectability}
\begin{tabular}{|lc|}
\hline
\textbf{Source} & \textbf{AI Probability (\%)} \\
\hline
Claude 3 Opus & 80.01 \\
Gemini 1.5 Flash & 71.57 \\
Claude 3 Sonnet & 69.53 \\
Llama 3.1 405B & 67.14 \\
Grok & 59.98 \\
GPT-3.5 & 58.49 \\
Llama3.1 8B & 51.53 \\
Gemini 1.5 Pro & 50.32 \\
GPT-4 & 42.50 \\
Students & 20.24 \\
Claude 3.5 Sonnet & 18.86 \\
Ground Truth & 3.28 \\
\hline
\end{tabular}
\end{table}

The analysis reveals a wide range of detectability scores across different generators. As expected, human-authored content in the Ground Truth condition received the lowest AI probability (3.28\%), consistent with its manual origin and natural linguistic variation. Similarly, student-generated content --despite being composed in guided instructional settings--- was assigned a relatively low probability (20.24\%), indicating that GPTZero still distinguishes it from LLM-produced content with reasonable confidence.
Interestingly, the output of Claude 3.5 Sonnet (18.86\%) was less detectable as AI-based than even the average student output. This suggests that this model generates \us that better mimic human linguistic variability, perhaps because it was one of the newest models in the cohort. 
Claude 3 Opus exhibits the highest detectability at 80.01\%, followed by Gemini 1.5 Flash and Claude 3 Sonnet. 

Newer LLMs tend to evade GPTZero more easily. Claude 3.5 Sonnet shows the lowest AI probability (18.86\%), nearly matching student text (20.24\%), and GPT‑4 (42.50\%) is less detectable than GPT‑3.5 (58.49\%). The trend is not absolute: LLaMA 3.1 405B scores 67.14\%, higher than the smaller 8B model (51.53\%), so scaling up does not always lower detectability. 
Note that lower detectability does not necessarily imply lower quality. Models that maximize consistency and syntactic correctness might reveal their artificial origin due to excessive regularity. Detectability is, therefore, best interpreted as a proxy for human likeness, not usefulness or correctness.

\begin{tcolorbox}[colback=gray!5!white, colframe=black!60!black, title=Insights on Detectability]
Detectability captures stylistic similarity to human writing, not necessarily its correctness or utility. Our analysis reveals that some advanced LLMs ---such as \textit{Claude 3.5 Sonnet} and \textit{GPT-4}--- produce \us that are less likely to be flagged as AI-generated, often aligning closely with student-authored content in human-likeness. This may stem from improved alignment and natural language variability.
However, trends are inconsistent: \textit{LLaMA 3.1 405B} is more detectable than its smaller sibling, illustrating that a larger size does not guarantee more human-like output. 
\end{tcolorbox}

\subsection{Answering RQ1}

Our analysis shows that LLMs can effectively generate \us that match the basic requirements represented in \gtus, achieving high coverage scores across models. This suggests that LLMs can align well with fundamental requirement patterns, often producing syntactically clean and semantically precise outputs. However, this high coverage may come at the cost of variability and originality. Compared to students, who display much higher output diversity, LLMs tend to generate more homogeneous stories, pointing to a narrower exploration of the requirement space. This lack of variability might limit their usefulness in early ideation phases where breadth is essential.
Detectability results add further nuance. Some recent and well-aligned models (e.g., Claude 3.5 Sonnet, GPT-4) produce \us that are not only high-quality but also closely mimic human writing in style, making them less likely to be flagged as AI-generated. However, this trend is inconsistent across all models or families, suggesting that factors beyond scale ---such as training data and decoding strategy--- play a critical role in shaping output human-likeness.

In summary, LLMs exhibit strong potential in supporting \us generation. However, their tendency toward structural regularity and limited variation differentiates them from human experts, who naturally infuse creativity and contextual nuance into requirements specification.

\begin{tcolorbox}[colback=gray!5!white, colframe=black!60!black, title=RQ1: Can LLMs generate \us similar to those created by human experts?]
LLMs can generate \us similar to human experts in terms of coverage and stylistic quality, particularly for basic requirements. However, they exhibit lower diversity and creativity, which may limit their effectiveness in exploratory or nuanced elicitation tasks. Recent models also show reduced detectability, indicating increased human-likeness in stylistic features.
\end{tcolorbox}

\section{LLM-based Assessment of User Stories}
\label{sec:assessment}
To understand whether LLMs can be used to evaluate the quality of \us (RQ2), we calculated the agreement between the labels generated by LLMs and those created by humans after conflict resolution. We used Cohen’s Kappa ($\kappa$)~\cite{mchugh2012interrater} metric, which quantifies the level of agreement between two raters while accounting for the agreement that could occur by chance.
A value of $\kappa$ of 1 indicates perfect agreement, 0 corresponds to agreement at the chance level, and negative values imply systematic disagreement. In our context, higher values of $\kappa$ indicate that the LLM evaluation behavior is more similar to human annotators, reflecting a better ability to evaluate the quality of \us.

\begin{table*}[t]
\centering
\caption{Inter-Annotator Agreement Before Conflict Resolution (Cohen's Kappa).}
\label{tab:human_agreement}
\begin{tabular}{lccc}
\hline
\textbf{Quality Criterion} & \textbf{Exp 1 vs Exp 2} & \textbf{Exp 1 vs Exp 3} & \textbf{Exp 2 vs Exp 3} \\
\hline
Feature Specificity (FS)     & Substantial (0.72)     & Substantial (0.79)     & Substantial (0.76) \\
Rationale Clarity (RC)       & Substantial (0.75)     & Substantial (0.67)     & Almost Perfect (0.81) \\
Problem Oriented (PO)        & Substantial (0.72)     & Substantial (0.72)     & Moderate (0.52) \\
Language Clarity (LC)        & Substantial (0.70)     & Almost Perfect (0.83)  & Substantial (0.79) \\
Internal Consistency (IC)    & Substantial (0.69)     & Almost Perfect (0.85)  & Substantial (0.72) \\
\hline
\textbf{Overall $\kappa$}    & Substantial (0.76)     & Substantial (0.78)     & Substantial (0.74) \\
\hline
\end{tabular}
\end{table*}

To focus on the most meaningful cases of disagreement, we adopted a conflict-aware formulation of $\kappa$. Specifically, a disagreement between an LLM and human annotation was only counted when there was a conflict (see Equation \ref{eq:conflict}). This mirrors the conflict resolution logic used in the manual annotation reconciliation process (as discussed in Section~\ref{sec:preparation}), where the key concern was whether a \us fundamentally fails to meet the acceptability criteria.
In this setting, $\kappa$ better reflects the alignment of an LLM with the decision boundaries that humans deem critical to determine \us adequacy.

Cohen’s Kappa was computed separately for each of the five quality dimensions: FS, RC, PO, LC, and IC (as defined in Section~\ref{sec:manuallabeling}). This allows us to identify which aspects of quality are more or less reliably assessed by different models.
In general, a higher $\kappa$ indicates a greater ability of the LLM to act as a reliable evaluator of \us quality, potentially reducing the effort required in manual validation workflows. In contrast, lower agreement highlights areas where LLMs struggle to interpret or apply quality criteria in a manner consistent with human experts, even when detailed codebooks are provided. Given the task, we configured all the LLMs with a temperature of $0$ (as suggested by Renze~\cite{renze2024effect}) to ensure deterministic behavior and eliminate randomness in generation, allowing for a more controlled and reproducible comparison across models. 

\subsection{Agreement between Human Experts}

To understand the results obtained by LLMs, here we show the agreement among human experts during the manual labeling phase and before the reconciliation process. 

The values reported in Table~\ref{tab:human_agreement} were interpreted according to standard qualitative thresholds for Cohen's Kappa: \textit{Poor} ($\kappa < 0$), \textit{Slight} ($0 \leq \kappa < 0.2$), \textit{Fair} ($0.2 \leq \kappa < 0.4$), \textit{Moderate} ($0.4 \leq \kappa < 0.6$), \textit{Substantial} ($0.6 \leq \kappa \leq 0.8$), and \textit{Almost Perfect} ($\kappa > 0.8$). 
The table shows that most of the comparisons yielded \textit{Substantial} agreement, with several criteria ---particularly Language Clarity and Internal Consistency--- reaching \textit{Almost Perfect} agreement in some pairs of annotators. Only one case fell in the \textit{Moderate} range, indicating a slightly lower consistency in judging the Problem-Oriented criterion between Expert 2 and Expert 3.
Overall, these results confirm a strong level of alignment among the annotators, although the task inherently involves a degree of subjectivity. Assessing the quality of \us requires interpretative judgment, especially when distinguishing between acceptable and good requirements or evaluating the rationale or conceptual clarity of \us. The observed levels of agreement demonstrate that, despite this subjectivity, the evaluation framework supported reasonably consistent scores among different human experts.

\begin{tcolorbox}[colback=gray!5!white, colframe=black!60!black, title=Insights on Human Agreement]
Human experts agreed strongly when identifying unacceptable \us. This suggests that despite the subjectivity of the task, clearly inadequate \us can be reliably detected using a structured evaluation process.
\end{tcolorbox}

\subsection{LLM-Human Agreement in \us Assessment}
\begin{figure*}[t] 
\centering 
\includegraphics[width=0.85\textwidth]{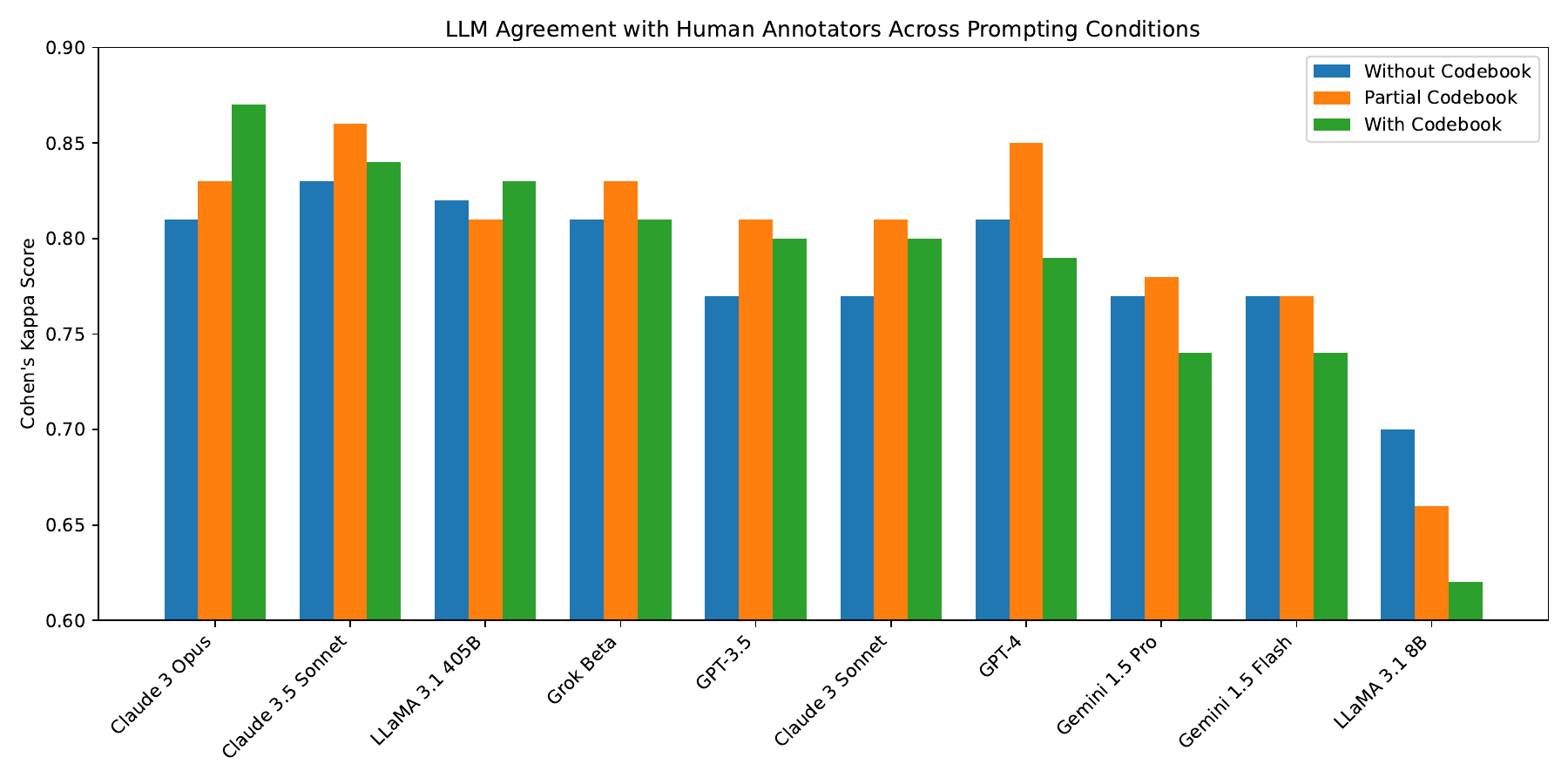}
\caption{Cohen’s Kappa between each LLM and the human-annotated ground truth (across all quality criteria).} 
\label{fig:llm_agreement_comparison} 
\end{figure*}

Figure~\ref{fig:llm_agreement_comparison} summarizes the evaluation of LLM labels and human-annotated ground truth under three prompting conditions: without codebook, with partial codebook, and with complete codebook (step \blackcircle{9}). Each condition represents an increasing level of guidance provided to the models, ranging from a minimal task description to detailed evaluation criteria. The figure shows Cohen’s Kappa scores for each LLM under these conditions, which captures how well their evaluations align with those of human experts.
In general, all models exhibit at least a \textit{Substantial} level of agreement across conditions, several reaching \textit{Almost Perfect} agreement ($\kappa > 0.8$) when provided with the complete codebook. This confirms that LLMs can mirror human decision making in evaluating the quality of \us, especially when provided with structured and explicit criteria. In some cases, such as Claude 3 Opus ($\kappa = 0.87$) and Claude 3.5 Sonnet ($\kappa = 0.84$), the agreement even exceeds the average inter-human reliability observed in our manual annotation phase, suggesting that LLMs can act as robust evaluators when adequately instructed.

The comparison across prompting conditions reveals a consistent trend: the codebook acts as an anchor that generally helps models interpret ambiguous or under-specified evaluation tasks. 
For example, Claude 3 Opus obtained $\kappa = 0.81$ without the codebook,  $\kappa = 0.83$ with partial codebook, and  $\kappa = 0.87$ with the full codebook.
Interestingly, while all models except LLaMa 3.1 8B and Gemeni 1.5 Flash benefit from the codebook, some show diminishing returns or slight reversals. For example, Claude 3.5 Sonnet and GPT-4 perform better with partial codebook guidance ($\kappa = 0.86/0.85$) than with the full version ($\kappa = 0.84/0.78$). This may indicate that excessive prompting or overly detailed instructions can sometimes interfere with the model's internal heuristics, introducing noise or unnecessary constraints.
Compared to human-human agreement levels, which were typically between $0.74$ and $0.78$, most top-performing LLMs match or exceed this threshold when appropriately guided. This alignment suggests that LLMs are not only reproducing surface-level patterns but are capable of modeling evaluative behavior comparable to that of domain experts. However, the extent of this capability depends on the clarity of the task framing, as demonstrated by the improvements across prompt conditions.

The lowest performing model in all conditions is LLaMA 3.1 8B, which reaches a maximum $\kappa$ of only $0.70$, still within the substantial range but trailing behind its larger counterpart LLaMA 3.1 405B (which reaches $0.83$). This reinforces the observation that model size and training scope play a role in the ability to generalize and follow abstract evaluative criteria. More powerful models exhibit better semantic understanding and are more sensitive to prompt structure.

The results show that LLMs are already competent evaluators of \us quality and can replicate human judgment to a high degree of fidelity. The provision of structured evaluation criteria significantly enhances their performance, confirming that prompt engineering and task framing are critical levers in deploying LLMs for quality assurance in requirements engineering. These findings position LLMs as promising tools for (partially) automating human-centered validation workflows.

\begin{table}[t]
\centering
\caption{LLM-Human Agreement (Cohen’s Kappa) by Quality Criterion – With Codebook.}
\label{tab:with_codebook_detailed}
\begin{tabular}{lccccc|c}
\hline
\textbf{LLM} & \textbf{FS} & \textbf{RC} & \textbf{PO} & \textbf{LC} & \textbf{IC} & \textbf{Overall} \\
\hline
Claude 3 Opus         & 0.89 & 0.86 & 0.72 & 0.92 & 0.97 & \textbf{0.87} \\
Claude 3.5 Sonnet     & 0.90 & 0.87 & 0.60 & 0.87 & 0.97 & \textbf{0.84} \\
LLaMA 3.1 405B        & 0.91 & 0.73 & 0.68 & 0.94 & 1.00 & \textbf{0.83} \\
Grok Beta             & 0.88 & 0.68 & 0.72 & 0.89 & 0.94 & \textbf{0.81} \\
Claude 3 Sonnet       & 0.86 & 0.64 & 0.68 & 0.90 & 0.95 & \textbf{0.80} \\
GPT-3.5               & 0.87 & 0.67 & 0.70 & 0.88 & 0.85 & \textbf{0.80} \\
GPT-4                 & 0.81 & 0.72 & 0.69 & 0.85 & 1.00 & \textbf{0.79} \\
Gemini 1.5 Pro        & 0.77 & 0.71 & 0.52 & 0.92 & 0.90 & \textbf{0.74} \\
Gemini 1.5 Flash      & 0.88 & 0.66 & 0.50 & 0.87 & 0.86 & \textbf{0.74} \\
LLaMA 3.1 8B          & 0.77 & 0.63 & 0.67 & 0.54 & 0.50 & \textbf{0.62} \\
\hline
\end{tabular}
\end{table}

Table~\ref{tab:with_codebook_detailed} provides a detailed view of the agreement between each LLM and the human-established ground truth under the most structured condition with the codebook prompt. The values represent $\kappa$ scores for each of the five quality criteria and an overall score calculated as the average across dimensions.
Across the board, most models achieve Substantial or Almost Perfect  agreement with human annotations, with Claude 3 Opus leading at an overall $\kappa$ of $0.87$. 
High-performing models tend to exhibit strong consistency across dimensions, with particularly strong agreement in LC and IC ---two relatively syntactic and less interpretative dimensions. Several models (e.g., GPT-4 and LLaMA 3.1 405B) achieve a perfect score ($\kappa = 1.0$) in Internal Consistency, likely due to the model’s capacity to recognize basic logical coherence within a short text.
More semantically nuanced criteria, such as RC and PO, show more variability across models. This aligns with the observation that these dimensions require deeper contextual understanding and often reflect implicit intent. For instance, Claude 3.5 Sonnet, while very strong on RC ($0.87$), drops to $0.60$ on PO, suggesting that it may prioritize clarity of purpose over critical assessment of solution bias.
In contrast, Gemini 1.5 Flash and LLaMA 3.1 8B fall behind, especially in LC and IC. The latter performs weakest overall ($\kappa = 0.62$), showing limited alignment across all five dimensions. These gaps may arise from a smaller model size (e.g., 8B vs. 405B), architectural limitations, or differences in alignment and fine-tuning processes.
While model size often correlates with stronger performance, this is not always the case. For example, GPT-3.5 performs as well as GPT-4 overall ($0.80$ vs. $0.79$) and is better in several individual dimensions. This suggests that alignment techniques and prompt compatibility may be as significant as scale in supporting high-quality assessment.

\begin{tcolorbox}[colback=gray!5!white, colframe=black!60!black, title=Insights on LLM-based \us Assessment]
Across all prompt conditions, most LLMs reached substantial to high agreement with human annotations, with several exceeding inter-human agreement levels. In the with-codebook condition, \textit{Claude 3 Opus} ($\kappa=0.87$), \textit{Claude 3.5 Sonnet} ($\kappa=0.84$), and \textit{LLaMA 3.1 405B} ($\kappa=0.83$) were the top performers. The agreement was strongest in syntactic dimensions (e.g., LC, IC), and more variable in semantically complex ones (e.g., RC, PO). Supplying the codebook consistently improved alignment across the models.
\end{tcolorbox}

\subsection{Answering RQ2}

The results presented in this section show that LLMs can act as effective evaluators of \us quality when provided with appropriate guidance. Using a conflict-aware Cohen’s Kappa metric, we measured the agreement between each LLM and the human-annotated ground truth. With structured prompts, especially when a detailed codebook is supplied, several LLMs achieved agreement scores on par with or even exceeding human-human agreement.

These findings suggest that LLMs can reliably identify \us that fail to meet basic acceptability thresholds, particularly in more syntactic dimensions like Language Clarity and Internal Consistency. However, performance remains more variable in semantically demanding areas such as Rationale Clarity and Problem Orientation, where domain understanding and implicit reasoning play a more significant role.
Thus, LLMs offer a viable support mechanism for quality assurance in requirements engineering, particularly for tasks that involve detecting basic structural and logical issues. However, their use should be calibrated with prompt design and accompanied by human oversight in more nuanced assessments.

\begin{tcolorbox}[colback=gray!5!white, colframe=black!60!black, title=RQ2: How can LLMs be used to evaluate the quality of \us?]
LLMs can reliably assess \us quality when equipped with clear evaluation criteria. They perform best on structurally grounded dimensions and can reduce human effort in large-scale assessments. However, their reliability depends on prompt structure, and human oversight remains important for semantically nuanced judgments. With proper setup, LLMs are promising tools for semi-automated quality assurance in requirements engineering.
\end{tcolorbox}

\section{Evaluating Quality of Generated User Stories}
\label{sec:quality}

We evaluated the syntactic and semantic qualities of \us generated by different LLMs to answer RQ3. 

To evaluate the syntactic quality of \us generated by the LLMs, we used the \textit{AQUSA}~\cite{lucassen2016improving} (\emph{Automatic Quality User
Story Artisan })  framework (step \blackcircle{10}). It applies a structured set of static checks based on a formal taxonomy of quality criteria, targeting syntactic, structural, and semantic correctness. The tool outputs detailed metrics that capture the presence and distribution of quality defects.

For each model, we report the total number of \us generated by each model (\textit{Total}), the number of \us that contained at least one defect (\textit{\# Def}), and the corresponding proportion of such \us over the total (\textit{\% Def}, computed as $\frac{\text{\# Def}}{\text{Total}} \times 100$). To assess how concentrated defects are within the faulty \us, we also compute the \textit{Average Defects} (\textit{Avg Def}), defined as the number of defects divided by the number of defected \us.

AQUSA categorizes defects into several distinct types, each representing a different class of syntactic quality issues. Among them:
\begin{itemize}
    \item \textit{atomic.conjunctions} (\textit{conj}): excessive or improper use of conjunctions (e.g., “and”) that compromise atomicity.
    \item \textit{uniform.uniform} (\textit{uniform}): inconsistencies in the format or structure of acceptance criteria across \us.
    \item \textit{minimal.brackets} (\textit{brackets}): incorrect use of brackets or parentheses.
    \item \textit{minimal.indicator\_repetition} (\textit{rep}): redundant specification of role or goal indicators.
    \item \textit{unique.identical} (\textit{identical}): repeated or duplicated acceptance criteria items that should be unique.
    \item \textit{well\_formed.no\_means} (\textit{no\_means}): missing specification of the ``means'' in the canonical form: ``As a \emph{role}, I want \emph{goal}, so that \emph{reason}''.
    \item \textit{well\_formed.no\_role} (\textit{no\_role}): omission of the user role.
    \item \textit{minimal.punctuation} (\textit{punct}): punctuation-related issues such as missing or duplicated symbols.
\end{itemize}

\noindent We reported all metrics as \% of the total number of \us, except for the absolute counts (Total, \# Defected) and averages.

To evaluate the semantic quality of \us, we performed a large-scale annotation experiment using the best-performing model of RQ2 (Claude 3 Opus) (step \blackcircle{11}). This model showed the highest agreement with human experts in structured assessment tasks, and therefore we selected it as a reference evaluator to label systematically \us generated by all LLMs, those written by students, and those present in the ground truth set.
Claude 3 Opus was prompted with the complete codebook and asked to assign scores to each \us according to the five predefined quality criteria.

As for the previous steps, each criterion was scored on a 3-point scale: 1 (non-acceptable), 2 (acceptable), or 3 (good). Based on these annotations, we computed two aggregate measures for each source.
\begin{itemize} \item \textit{Rejection Rate (\%)}: the percentage of \us that received a score of 1 (non-acceptable) on at least one of the five quality criteria. This metric reflects how many \us fail to meet minimal quality expectations. 
\item \textit{Avg Score $[1,3]$}: the average of all quality scores assigned across all criteria and \us from a given source. This provides an overall indicator of the typical quality level observed.
\end{itemize}
This setup allowed us to compare the perceived quality of \us across different generators using a consistent, automated, and fine-grained framework, allowing qualitative comparisons of both quantitative and per-criterion qualitative comparisons. 

\subsection{Quantitative Evaluation}

\begin{table*}[t]
\centering
\caption{All Defect Metrics in a Single Table (sorted by \% Defected).}
\label{tab:all-in-one}
\begin{tabular}{lcccccccccccc}
\hline
\textbf{Model} & \textbf{Total} & \% \textbf{Def} & \# \textbf{Def} & \textbf{Avg Def} & \textbf{conj} & \textbf{uniform} & \textbf{brackets} & \textbf{rep} & \textbf{identical} & \textbf{no\_means} & \textbf{no\_role} & \textbf{punct} \\
\hline
Claude Sonnet 3.5 & 1500 &  2.20 & 33   & 1.00 &  1.67 &  0.47 & 0.00 & 0.00 & 0.07 & 0.00 & 0.00 & 0.00 \\
GPT-3.5 Turbo     & 1360 & 16.69 & 232  & 1.02 & 13.68 &  2.79 & 0.00 & 0.00 & 0.59 & 0.00 & 0.00 & 0.00 \\
Gemini 1.5 Pro    & 1595 & 18.18 & 326  & 1.12 &  5.27 & 12.66 & 1.57 & 0.44 & 0.19 & 0.31 & 0.00 & 0.07 \\
GPT-4             & 1490 & 21.14 & 378  & 1.20 & 10.87 & 11.01 & 0.00 & 0.27 & 1.21 & 2.01 & 0.00 & 0.00 \\
Gemini 1.5 Flash  & 1506 & 23.31 & 358  & 1.02 & 21.05 &  1.59 & 0.80 & 0.27 & 0.07 & 0.00 & 0.00 & 0.00 \\
Llama 3.1 405B     & 1452 & 30.37 & 579  & 1.31 & 24.24 &  0.07 & 0.00 & 0.00 & 7.02 & 4.27 & 4.27 & 0.00 \\
Claude Opus 3     & 1481 & 38.28 & 641  & 1.13 & 28.22 & 13.91 & 0.95 & 0.14 & 0.00 & 0.00 & 0.00 & 0.00 \\
Llama 3.1 8B      & 1460 & 41.58 & 793  & 1.31 & 29.38 &  1.10 & 0.14 & 0.14 & 15.14 & 4.25 & 4.11 & 0.00 \\
Grok Beta         & 1500 & 51.47 & 1525 & 1.98 & 22.40 & 19.13 & 0.20 & 0.60 & 23.00 & 26.27 & 10.07 & 0.07 \\
Claude Sonnet 3   & 1196 & 55.94 & 798  & 1.19 & 57.78 &  8.03 & 0.75 & 0.17 & 0.00 & 0.00 & 0.00 & 0.00 \\
\hline
\end{tabular}
\end{table*}

Table~\ref{tab:all-in-one} presents a comprehensive view of the AQUSA-based defect analysis in all LLMs evaluated. The results indicate substantial differences in quality across models in terms of the quantity and nature of the defects identified.

Claude Sonnet 3.5 stands out with a remarkably low defect rate, with only 2.20\% of \us that contain any issue and 33 defects over 1500 \us. This reflects a high level of structural and syntactic correctness, further supported by its low average defect count per defective \us (1.00). On the opposite end of the spectrum, Claude Sonnet 3 exhibits the highest defect rate at 55.94\%, suggesting a significant drop in quality relative to its updated counterpart. In particular, this earlier variant shows a very high percentage of atomic conjunction issues (57.78\%), which alone may explain much of the inflated defect rate.

Grok Beta also shows critical quality concerns. Despite generating the same number of \us as Claude Sonnet 3.5 (1500), it produces the highest total number of defects (1525) and a high average of 1.98 defects per defective \us. These defects are frequent and structurally severe. Grok exhibits elevated rates of \texttt{unique.identical} (23.00\%), \texttt{well\_formed.no\_means} (26.27\%), and \texttt{well\_formed.no\_role} (10.07\%), pointing to systemic issues in the composition of \us.
Looking at large models, LLaMA 3.1 405B shows a moderate defect rate (30. 37\%), with notable issues related to identical acceptance criteria (7.02\%) and missing components (\texttt{no\_means} and \texttt{no\_role} at 4.27\% each). Despite its size, it performs worse than some smaller models, such as Gemini 1.5 Flash, which achieves a lower defect rate (23.31\%). This suggests that size alone is not a sufficient indicator of reliability in the generation of \us.
GPT-3.5 Turbo and Gemini 1.5 Pro exhibit balanced behavior with relatively low \% Defected (16.69\% and 18.18\%, respectively) and average defect counts close to 1. These models suffer from milder issues, such as inconsistent formatting (\texttt{uniform}) and atomic conjunctions.  GPT-4, although a more capable model in general NLP tasks, shows a higher defect rate (21.14\%) than GPT-3.5 Turbo, though with moderate per-\us defect density (1.20).

Another key trend concerns the nature of defects. Atomic conjunctions (\texttt{conj}) are the most frequently reported issue overall, representing more than 57\% in some models. This suggests that while models often generate syntactically correct \us, they tend to concatenate multiple user intents, compromising atomicity, a known challenge in \us authoring. Models such as LLaMA 3.1 8B and Claude Opus 3 also exhibit high conjunction rates (29.38\% and 28.22\%, respectively), confirming that this is a widespread defect class even among top performing systems.


\begin{tcolorbox}[colback=gray!5!white, colframe=black!60!black, title=Insights on Quantitative Evaluation of Generated \us]
Quantitative analysis with AQUSA reveals substantial variation in quality across LLMs. \textit{Claude 3.5 Sonnet} shows the lowest defect rate (2.20\%), while older or less aligned models like \textit{Claude Sonnet 3} and \textit{Grok Beta} produce many more issues, especially related to atomicity and missing structure. The size of the model alone does not guarantee better performance ---\textit{LLaMA 3.1 405B} underperforms smaller peers. Common defects, such as excessive conjunctions, are widespread, even among advanced models.
\end{tcolorbox}

\subsection{Qualitative Evaluation}

\begin{table*}[t]
\centering
\caption{Qualitative Evaluation by Claude 3 Opus: Rejection Rate, Average Score, and Per-Criterion Ratings.}
\label{tab:qualitative_results}
\begin{tabular}{lccccccc}
\hline
\textbf{LLM} & \textbf{Rejection \%} & \textbf{Avg Score} & \textbf{FS} & \textbf{RC} & \textbf{PO} & \textbf{LC} & \textbf{IC} \\
\hline
Ground Truth      & 9.43  & 2.78 & 2.74 & 2.66 & 2.85 & 2.68 & 2.98 \\
Students          & 14.64 & 2.70 & 2.54 & 2.68 & 2.68 & 2.67 & 2.92 \\
Claude 3 Opus     & 16.55 & 2.78 & 2.60 & 2.71 & 2.71 & 2.92 & 2.99 \\
LLaMA 3.1 405B    & 19.23 & 2.81 & 2.74 & 2.65 & 2.81 & 2.88 & 2.98 \\
Gemini 1.5 Pro    & 20.33 & 2.77 & 2.63 & 2.58 & 2.81 & 2.85 & 2.99 \\
LLaMA 3.1 8B      & 22.10 & 2.76 & 2.78 & 2.39 & 2.83 & 2.87 & 2.94 \\
Gemini 1.5 Flash  & 24.85 & 2.81 & 2.80 & 2.36 & 2.94 & 2.93 & 3.00 \\
Grok Beta         & 25.70 & 2.70 & 2.71 & 2.13 & 2.82 & 2.88 & 2.98 \\
Claude 3.5 Sonnet & 33.24 & 2.76 & 2.78 & 2.22 & 2.96 & 2.88 & 2.98 \\
Claude 3 Sonnet   & 48.37 & 2.61 & 2.56 & 1.95 & 2.70 & 2.83 & 2.99 \\
GPT-4             & 63.47 & 2.48 & 2.48 & 1.54 & 2.75 & 2.71 & 2.91 \\
GPT-3.5           & 77.34 & 2.23 & 2.15 & 1.25 & 2.51 & 2.47 & 2.77 \\
\hline
\end{tabular}
\end{table*}

Table~\ref{tab:qualitative_results} presents the measured quality of \us produced by various LLMs, students, and the ground truth set, as evaluated by Claude 3 Opus. The evaluation reveals several trends, particularly regarding the rejection rate, the average score, and dimension-specific strengths and weaknesses.

\gtus achieved the best performance overall, with the lowest rejection rate (9.43\%) and one of the highest average scores (2.78). This confirms the quality of these manually curated \us, which serve as the upper bound in our comparison. \stuus followed closely, with a rejection rate of 14.64\% and an average score of 2.70. This suggests that, despite being non-experts, students can produce reasonably high-quality requirements when guided by instructional material and structure.
Among LLMs, Claude 3 Opus, the model used as the evaluator, scored its generated \us very high, with an average score of 2.78 and a rejection rate of 16. 55\%, closely aligning with human-created content. Some LLMs outperformed Claude 3 Opus in terms of average score: both LLaMA 3.1 405B and Gemini 1.5 Flash reached the highest average score of 2.81. However, this came at the cost of higher rejection rates (19.23\% and 24.85\%, respectively). This indicates that while these LLMs can produce highly rated \us, they are also more prone to generating \us that do not meet the minimum quality standards.

A more detailed view emerges when considering the per-criterion scores. Almost all models perform well in IC and LC, with values often approaching or even reaching 3.00. This is expected, as these dimensions are highly dependent on surface-level linguistic patterns, coherence, and grammar, areas in which LLMs excel. 
In contrast, models perform significantly worse in RC, reflecting the ability to explain the ``why'' behind the requirement. This is where human-created content has a noticeable advantage. \gtus and \stuus achieve RC scores of 2.66 and 2.68, respectively, while most LLMs lag behind. For example, GPT-4 scores only 1.54 and GPT-3.5 only 1.25, suggesting a systematic difficulty in making explicit the underlying rationale, even when the rest of the \us is structurally sound.
GPT-3.5 has the highest rejection rate (77.34\%) and the lowest average score (2.23), with a lower performance in RC (1.25) and LC (2.47). GPT-4, while stronger overall, also suffers from a high rejection rate (63.47\%) and similarly low RC (1.54). These results suggest that, despite their strong performance in many benchmarks, these models may be less reliable in structured requirements tasks without domain adaptation.
Interestingly, even among stronger models like Claude 3.5 Sonnet, the rejection rate remains high (33.24\%) despite a solid average score (2.76). This again points to inconsistency: some \us are rated very well, while others fall below acceptable thresholds.

\begin{tcolorbox}[colback=gray!5!white, colframe=black!60!black, title=Insights on Qualitative Evaluation of Generated \us]
LLMs can generate \us of competitive quality in dimensions such as LC and IC, where scores often match or exceed human performance. However, deeper aspects such as RC remain challenging, with most models lagging behind students and ground truth. Human-generated \us remain more consistently acceptable, reinforcing the need for human oversight.
\end{tcolorbox}

\subsection{Answering RQ3}

The combined evaluation reveals a clear picture of LLM performance in generation of \us. The syntactic and structural quality is generally strong, especially in terms of linguistic clarity and consistency. However, models often struggle with deeper semantic aspects, such as articulating the rationale behind features, where human-written \us retain a clear advantage. Quality variability remains a concern: some LLMs produce excellent outputs but lack consistency, with a non-negligible share of unacceptable \us. These fluctuations highlight the importance of controlled deployment and the need for human-in-the-loop validation. Model size or recency does not uniformly predict quality: training objectives and alignment matter equally, if not more. Overall, LLMs are promising requirements specification tools, but their limitations call for careful integration and monitoring in practical workflows.

\begin{tcolorbox}[colback=gray!5!white, colframe=black!60!black, title=RQ3: What is the quality of \us generated by LLMs?]
LLMs can generate high-quality \us, especially in terms of clarity and structure. However, they still fall short of the aspects that require reasoning and justification. Quality varies significantly across and within models, with no guarantee that model scale ensures the generation of high quality \us. This finding suggests that   human oversight is still necessary.
\end{tcolorbox}

\section{Threats to Validity}
\label{sec:threats}

In the following, we discuss the main threats that could impact the validity of our results~\cite{DBLP:books/daglib/p/WohlinHH06}.

\textbf{Internal Validity.} The outputs of LLMs are inherently non-deterministic and may vary between runs. To mitigate this, we generated 30 independent instances of user stories for each LLM model, reducing the impact of random fluctuations in individual outputs. Furthermore, when employing LLMs for the evaluation of user stories, we set the generation temperature at 0, following best practices~\cite{renze2024effect}, to ensure deterministic behavior and reproducibility. 
Regarding the qualitative evaluation, given the large scale of the dataset (around 14,000 user stories), we automated the annotation process using Claude 3 Opus, the best performing model from our earlier evaluation. This choice could introduce a bias in favor of Claude 3 Opus when it evaluates its own generated user stories. However, it is important to emphasize that the objective of this study is not to declare the absolute best-performing model, but rather to investigate the broader potential and limitations of LLMs in supporting the requirements elicitation process. Although there may be slight biases in the evaluation of specific models, overall findings regarding LLM capabilities and weaknesses remain robust and are not substantially affected by this choice.

Another potential internal threat to validity concerns the possibility that LLMs may have been exposed to the \gtus dataset during training. The nature of LLMs inherently reduces the risk of direct memorization of small datasets since these models rely on statistical patterns learned from massive-scale data and typically do not memorize exact instances of small, individual datasets~\cite{carlini2023quantifying}. Their predictions are based on probabilistic reasoning over generalized linguistic structures rather than specific textual content, which significantly limits verbatim recall of minor datasets like \gtus.

\textbf{External Validity.} Our study is carried out within the context of a single application domain, namely, a camping management system, which may limit the generalizability of the results to other domains. Requirements in different domains (e.g., finance, healthcare, embedded systems) may exhibit different linguistic structures, abstraction levels, or domain-specific terminology that could affect LLM performance. However, we deliberately selected a realistic and sufficiently complex system with a ground truth set of user stories manually curated by domain experts. This choice ensures that the evaluation is grounded in a concrete and representative scenario. Moreover, the methodology and evaluation pipeline we propose (e.g., coverage, diversity, defect analysis, qualitative scoring) are domain-agnostic and can be reused in future studies to assess generalizability across multiple contexts. 

\textbf{Construct Validity.} Construct validity concerns whether the evaluation metrics and procedures accurately capture the intended concepts. In our study, we rely on well-established frameworks: AQUSA for syntactic quality and the Quality User Story framework for qualitative assessment. Although these frameworks are grounded in literature and practice, the application of their criteria, particularly in manual and model-based labeling, can be inherently subjective. To reduce interpretation variability, we developed a detailed evaluation codebook that was iteratively refined and reviewed by all authors. 

\section{Related Work}
\label{sec:related}

\subsection{LLMs in Requirements Engineering}

LLMs have been used extensively in software engineering, particularly to support code generation, testing, and maintenance~\cite{Fan.2023}. 
LLMs can help generate complete software requirements specifications significantly faster~\cite{Ozkaya.IEEESoftware.2023} and, more generally, can improve the efficiency and accuracy of requirements-related tasks~\cite{Arora.2024}. Despite their potential in supporting requirements engineering activities (e.g., modeling with different formalisms and domains~\cite{Chen2023GPT4GoalModels, Chen2023AutomatedDomainModeling, deKinderen2024LegalGRL, Camara2023GenerativeAIUML} or traceability~\cite{Rodriguez:2023, North:2024, Wei:2024}) and the initial exploration of their effectiveness in addressing software engineering problems~\cite{hemmat2025research, cheng2025generativeairequirementsengineering, marques2024chatgpt}, LLMs have not been widely applied to support requirements engineering activities. 

Previous work has focused on using LLMs to reduce requirements ambiguity and support requirements elicitation. For example, Luitel et al.~\cite{Luitel.RE.2024} propose an approach to reduce omissions in software requirements by using BERT~\cite{devlin2019bert} to predict missing keywords and concepts. Fantechi et al.~\cite{Fantechi.RE.2023} measure the performance of ChatGPT in finding inconsistencies in requirements. They notice that LLMs cannot substitute expert judgment but could rather complement manual analysis and speed up human annotation. White et al.~\cite{White.2024} present a set of prompt patterns aimed at evaluating completeness (i.e., whether the specification covers relevant system features) and ambiguity of a requirement specification. Similarly, Lubos et al.~\cite{lubos2024leveraging} use LLMs to detect quality issues in software requirements, such as ambiguity, incompleteness, and other categories
defined in ISO 29148, highlighting the ability of LLMs to enhance requirements quality assurance processes.

G{\"o}rer et al.~\cite{Gorer.REW.2023} use LLMs (ChatGPT and Bard) to generate interview scripts for requirements elicitation. The authors explore an iterative approach in which the LLM is successively presented with
initial segments of the interview script, enabling the step-by-step
generation of subsequent content. This iterative approach allows the LLM to avoid common mistakes made by analysts during stakeholder interviews (e.g., asking long or irrelevant questions)~\cite{Bano.REJ.2019}. 
Elicitron~\cite{Elicitron} uses LLMs to simulate a diverse set of user agents and automate requirements elicitation interviews. The approach proposes a context-aware generation
method that maximizes the diversity of requirements. It also allows for the creation of user agents that can support the identification of latent needs (i.e., unarticulated and unexpected factors that strongly influence product desirability). 
We focus on using LLMs to automate the generation of user stories from the IBL process. To achieve this aim, we use LLMs to emulate the IBL process conducted by students in previous work~\cite{Ferrari.ICSE.2023}. Unlike Elicitron~\cite{Elicitron}, we do not explore the ability of LLMs to diversify user agents. 

Recent works have explored the use of LLMs to bridge the gap between informal natural language requirements and formal specifications. For example, Guidotti et al.~\cite{guidotti2024translating} introduced an approach that leverages LLMs to translate natural language requirements into Property Specification Patterns, facilitating the formalization process for domain experts without deep expertise in formal methods. Similarly, Cosler et al.~\cite{cosler2023nl2spec} developed nl2spec, a framework that interactively translates unstructured natural language into temporal logic specifications using LLMs, enhancing the accessibility of formal verification. Reinpold et al.~\cite{reinpold2024verifying} investigated the potential of LLMs in verifying technical system specifications, demonstrating that models such as GPT-4o and Claude 3.5 Sonnet can effectively assess the fulfillment of requirements in system specifications. 
A general overview of the application of LLMs to promote the usability of formal requirements has been presented in \cite{ferrari2025formal}, which proposes a roadmap for the use of LLMs to assist in understanding and generating formal requirements. This work highlights the promising role of LLMs in making formal requirements engineering more approachable and efficient.

Other work leveraged deep learning models and LLMs to automate the elicitation of software requirements. For example, Gudaparthi et al.~\cite{Gudaparthi.RE.2023} generate adversarial examples to identify candidates for novel requirements. They do so by applying small changes (perturbations) to descriptions of original requirements using a deep neural network. The generated requirements are traceable to existing software features, ensuring transparency and facilitating discussions among requirements engineers and developers. Bencheikh and H{\"o}glund~\cite{Bencheikh.2023} studied how ChatGPT-generated requirements align with human-written requirements. Humans and ChatGPT were provided with a domain description and requested to generate seven functional and two non-functional requirements, without receiving additional contextual information of how requirements should be specified.
ChatGPT demonstrated remarkable time efficiency compared to human participants without significantly compromising the quality of generated requirements. Similarly, Ronanki et al.~\cite{Ronanki.SEAA.2023} formulated six questions to elicit requirements using ChatGPT. Using the same six questions, the authors obtained requirements from five RE experts from academia and industry through interview-based surveys. Compared to the requirements generated by RE experts, ChatGPT-generated requirements were more abstract, atomic, consistent, correct, and understandable.
Krishna et al.~\cite{krishna2024using} conducted an empirical study comparing GPT-4 and CodeLlama with entry-level engineers to produce software requirements specifications, finding that LLMs can produce SRS drafts of comparable quality to those created by novice engineers, with GPT-4 demonstrating a greater ability to identify and rectify issues within requirements documents. 
Nouri et al.~\cite{Nouri.RE.2024} apply LLMs to perform safety analysis of the requirements of autonomous vehicles. More precisely, LLMs are used to identify hazardous events arising from a specific functionality, and elicit safety goals associated with a specific function and relevant safety goals associated with a hazardous event. The use of LLMs in this context allowed for complete safety analysis in a more efficient and effective way.
Unlike this work, we focus on generating requirements expressed as user stories and assessing the effectiveness of LLMs in evaluating the quality of the generated user stories.

\subsection{User Stories Evaluation}
The Quality User Story (QUS) framework~\cite{Lucassen.RE.2015} is a collection of 13 criteria that determine the quality of user stories in terms of syntax (well-formed, atomic, minimal), pragmatics (full sentence, estimatable, unique, uniform, independent and complete), and semantics (conceptually sound, problem-oriented, unambiguous, conflict-free).  Lucassen et al.~\cite{Lucassen.REJ.2016} also proposed an automated tool (AQUSA) to automatically evaluate the syntactical and pragmatic qualities of user stories using Stanford CoreNLP and the Natural Language Toolkit. The QUS framework and related toolkit have been extensively used in previous work (e.g.,~\cite{Lucassen.REFSQ.2017,Wautelet.2019,Wouters.REJ.2022}) to evaluate the quality of user stories. 
In this paper,  we focus on the use of LLMs to automatically assess the semantic qualities of user stories, which has not been supported in previous work~\cite{Lucassen.REJ.2016}.

\section{Conclusions}
\label{sec:conclusions}
This study shows that LLMs can support requirements engineering by generating \us with high syntactic and stylistic quality and by reliably assessing their semantic quality when guided by structured criteria. LLM-generated \us achieved higher coverage than those written by students but showed lower diversity and creativity, limiting their value in exploratory contexts. While newer models produce more human-like \us, they still meet acceptance criteria less consistently than human-authored ones. For assessment, several LLMs matched or exceeded inter-human agreement when using a detailed codebook. Overall, LLMs offer useful support for semi-automated elicitation and quality assessment, but prompt design and human oversight remain essential. Future work will investigate reasoning models and assess their capability to both generate and evaluate \us.

\section*{Data Availability}

We provide a complete replication package that contains all the code, datasets, the codebook, and evaluation scripts used in this study. The materials are publicly available at: \url{https://zenodo.org/records/15240143}.

\bibliographystyle{IEEEtran}
\bibliography{main}

\end{document}